%% file: cp_review_arxiv.tex
\documentclass[]{arxiv}

\usepackage{epstopdf}
\usepackage[caption=false]{subfig}

\usepackage[numbers,sort&compress]{natbib}
\bibpunct[, ]{[}{]}{,}{n}{,}{,}
\makeatletter
\def\NAT@def@citea{\def\@citea{\NAT@separator}}
\makeatother

\theoremstyle{plain}

\theoremstyle{definition}

\theoremstyle{remark}

\usepackage{amsmath}
\usepackage{braket}
\usepackage{tikz}

\usepackage{pgfplots} 
\usepackage{filecontents}			
\usetikzlibrary{arrows.meta,arrows,shapes,backgrounds,fit,decorations.pathreplacing,chains,snakes,positioning,angles,quotes,decorations.pathmorphing}  
\usepgfmodule{nonlineartransformations}
\usepackage{subfig}
\usepackage{xcolor}
\usepackage[raggedrightboxes]{ragged2e}
\usepackage{graphicx}
\usepackage{soul}
\usepackage{textcomp}
\DeclareRobustCommand\openone{\leavevmode\hbox{\small1\normalsize\kern-.33em1}}
\DeclareRobustCommand\openonesmall{\leavevmode\hbox{\footnotesize1\small\kern-.30em1}}

\newcommand{\cnot}{\leavevmode\hbox{\footnotesize{CNOT }}}
\newcommand{\cnoteq}{\leavevmode\hbox{\footnotesize{CNOT}}}

\begin{document}


\title{Quantum Error Correction: An Introductory Guide}

\author{
\name{Joschka Roffe\thanks{CONTACT Joschka Roffe. Email: j.roffe@sheffield.ac.uk}}
\affil{Department of Physics \& Astronomy, University of Sheffield, Sheffield, S3 7RH, United Kingdom}
}

\maketitle

\begin{abstract}
Quantum error correction protocols will play a central role in the realisation of quantum computing; the choice of error correction code will influence the full quantum computing stack, from the layout of qubits at the physical level to gate compilation strategies at the software level. As such, familiarity with quantum coding is an essential prerequisite for the understanding of current and future quantum computing architectures. In this review, we provide an introductory guide to the theory and implementation of quantum error correction codes. Where possible, fundamental concepts are described using the simplest examples of detection and correction codes, the working of which can be verified by hand. We outline the construction and operation of the surface code, the most widely pursued error correction protocol for experiment. Finally, we discuss issues that arise in the practical implementation of the surface code and other quantum error correction codes.

\end{abstract}


\section{Introduction}

In place of the bits in traditional computers, quantum computers work by controlling and manipulating quantum bits (qubits). Through the precise control of quantum phenomena such as entanglement, it is in principle possible for such qubit-based devices to outperform their classical counterparts. To this end, efficient quantum computing algorithms have been developed with applications such as integer factorisation \cite{Shor97}, search \cite{Grover96}, optimisation \cite{Schuld14} and quantum chemistry \cite{Aspuru-Guzik05}.

There is currently no preferred qubit technology; a variety of physical systems are being explored for use as qubits, including photons \cite{Wang2016,Qiang18}, trapped ions \cite{Randall15,Ballance16,Brandl16,Debnath16}, superconducting circuits \cite{Chow12,Chen2014,Wendin2017} and spins in semiconductors \cite{Kane1998,Hill2015,vanderHeijden18}. A shortcoming shared by all of these approaches is that it is difficult to sufficiently isolate the qubits from the effects of external noise, meaning errors during quantum computation are inevitable. In contrast, bits in a classical computer are typically realised as the robust \textit{on/off} states of transistor switches which are differentiated by billions of electrons. This provides classical bits with high error margins that near-eradicate failures at the physical level. For quantum computers, where qubits are realised as fragile quantum systems, there is no such security against errors. As such, any circuit-model quantum computer based on current and future qubit technologies will require some sort of active error correction.

Driven by the demands of high-performance communication networks and the Internet, there is a well-developed theory of classical error correction \cite{Shannon49,Hamming50,MacKay03a}. However, adapting existing classical methods for quantum error correction is not straightforward. Qubits are subject to the no-cloning theorem \cite{Wootters82}, meaning quantum information cannot be duplicated in the same way as classical information. Furthermore, it is not possible to perform arbitrary measurements on a qubit register due to the problem of wavefunction collapse. It was initially feared that these constraints would pose an insurmountable challenge to the viability of quantum computing. However, a breakthrough was reached in 1995 by Peter Shor with a paper proposing the first quantum error correction scheme \cite{Shor95}. Shor's method demonstrated how quantum information can be redundantly encoded by entangling it across an expanded system of qubits. Subsequent results then demonstrated that extensions to this technique can in principle be used to arbitrarily suppress the quantum error rate, provided certain physical conditions on the qubits themselves are met \cite{Preskill385,Kitaev1997,Aharonov:1997:FQC:258533.258579,Knill1998,Gottesman98}. It was with these developments in quantum error correction that the field of quantum computing moved from a theoretical curiosity to a practical possibility.

Many reviews have been written covering quantum error correction and its associated subfields \cite{Gaitan08,Gottesman09,Djordjevic12,Devitt13,Lidar13,Terhal15,Campbell17}. This work is intended as an introductory guide where we describe the essential concepts behind quantum error correction codes through the use of simple examples. The ultimate aim is to provide the reader with sufficient background to understand the construction and operating principles behind the surface code, the most widely pursued error correction scheme for experimental implementation \cite{Fowler12}. Crucially, our descriptions of the surface code do not rely upon terminology from topology and homology, as is the case with many of the original sources. Whilst this review does not require prior knowledge of coding theory or error correction, we do assume an understanding of elementary quantum mechanics and the circuit model of quantum computing. The reader should be comfortable with quantum circuit notation, as seen for example in \cite{Nielsen2010}, and be familiar with standard gates such as the Hadamard gate ($H$), the controlled-{NOT} gate (\cnoteq) and measurement operations in the computational basis. A brief outline of these gates, as well as the conventions we adopt for labelling quantum states and operators, can be found in appendices A-C.

In section \ref{sec:class_to_quantum}, we begin by explaining the differences between bits and qubits, before describing the principal challenges in designing quantum error correction codes. Section \ref{sec:2qubit} outlines how quantum information is redundantly encoded, and explains how errors can be detected by performing projective measurements. In section \ref{sec:stab_codes}, we introduce the stabilizer framework which allows for the construction a large class of quantum error correction codes. Following this, the surface code is described in section \ref{sec:surface}. Finally, in section \ref{sec:practical}, we discuss some of the practical issues that arise when considering the implementation of quantum error correction codes on realistic hardware.

\section{From classical to quantum error correction} \label{sec:class_to_quantum}

Classical information technologies employ binary encodings in which data is represented as sequences of bits takings values  `$0$' or `$1$'. The basic principle behind error correction is that the number of bits used to encode a given amount of information is increased. The exact way in which this \textit{redundant} encoding is achieved is specified by a set of instructions known as an \textit{error correction code} \cite{Hamming50,MacKay03a}.

The simplest example of an error correction code is the three-bit repetition code, the encoder for which duplicates each bit value $0\rightarrow 000$ and $1\rightarrow111$. More formally, we can define the three-bit encoder as a mapping from a `raw' binary alphabet $\mathcal{B}$ to a code alphabet $C_3$
\begin{equation}
\mathcal{B}=\{0,1\} \xrightarrow{three-bit \ encoding} C_{\rm 3} =\{000,111\}\rm,
\end{equation}        
where the encoded bit-strings `$000$' and `$111$' are referred to as the \textit{logical codewords} of the code $C_3$. As an example, consider the simple case where we wish to communicate a single-bit message `$0$' to a recipient in a different location. Using the three bit encoding, the message that we would send would be the `$000$' codeword.

Now, imagine that the message is subject to a single bit-flip error during transmission so that the bit-string the recipient receives is `$010$'. In this scenario, the recipient will be able to infer that the intended codeword is `$000$' via a majority vote. The same will be true for all cases where the codeword is subject to only a single error. However, if the codeword is subject to two bit-flip errors, the majority vote will lead to the incorrect codeword. The final scenario to consider is when all three bits are flipped so that the codeword `$000$' becomes `$111$'. In this case, the corrupted message is also a codeword: the recipient will therefore have no way of knowing an error has occurred. The \textit{distance} of a code is defined as the minimum number of errors that will change one codeword to another in this way. We can relate the distance $d$ of a code to the number of errors it can correct as follows
\begin{equation}
d=2t+1\rm.
\end{equation}
where $t$ is the number of errors the code can correct. It is clear that the above equation is satisfied for the three-bit code where $t=1$ and $d=3$.

In general, error correction codes are described in terms of the $[n,k,d]$ notation, where $n$ is the total number of bits per codeword, $k$ is the number of encoded bits (the length of the original bit-string) and $d$ is the code distance. Under this notation, the three-bit repetition code is labelled $[3,1,3]$.

\subsection{From bits to qubits}

In place of bits in classical systems, the fundamental unit of quantum information is the \textit{qubit}. The general qubit state can be written as follows
\begin{equation}\label{eq:qubit_gen_state}
\ket{\psi} = \alpha \ket{0} + \beta\ket{1}\rm,
\end{equation}
where $\alpha$ and $\beta$ are complex numbers that satisfy the condition $|\alpha |^2 + |\beta|^2 =1$. Details regarding the notation we use to represent quantum states can be found in appendix \ref{sec:notation}. Qubits can encode information in a superposition of their basis states, meaning quantum computers have access to a computational space that scales as $2^n$ where $n$ is the total number of qubits \cite{Nielsen2010}. It is by exploiting superposition, in combination with other quantum effects such as entanglement, that it is possible to construct algorithms that provide a quantum advantage \cite{Shor97,Grover96}. However, if such algorithms are ever to be realised on current or future quantum hardware, it will be necessary for the qubits to be error corrected.

\subsection{The digitisation of quantum errors}\label{sec:digitisation}

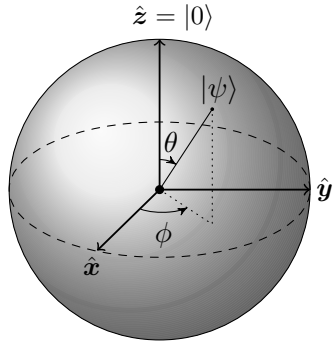
\begin{figure}
	\centering
	\input{bloch_sphere/bloch.tikz}
	\caption{In the geometric representation, the state of a qubit $\ket{\psi}=\cos{\frac{\theta}{2}}\ket{0}+e^{i\phi}\sin{\frac{\theta}{2}}\ket{1}$ can be represented as a point on the surface of a Bloch sphere.}
	\label{fig:bloch}
\end{figure}

In classical information, bits are either in the `0' or `1' state. Therefore, the only error-type to be considered is the bit-flip that takes $0\rightarrow1$ and vice-versa. In contrast, the general qubit state defined in equation (\ref{eq:qubit_gen_state}) can assume a continuum of values between its basis states. From the perspective of developing error correction codes, this property is problematic as it means the qubit is subject to an infinite number of errors. To illustrate this more clearly, it is useful to rewrite the general qubit state in terms of a geometric representation given by
\begin{equation}
\ket{\psi}=\cos{\frac{\theta}{2}}\ket{0}+e^{i\phi}\sin{\frac{\theta}{2}}\ket{1}\rm,
\end{equation}
where the probability amplitudes maintain the condition that $|\cos{\frac{\theta}{2}}|^2+| e^{i\phi}\sin{\frac{\theta}{2}}|^2=1$. In this form, the qubit state corresponds to a point, specified by the angles $\theta$ and $\phi$, on the surface of a so-called Bloch sphere. An example state in this representation is shown in figure \ref{fig:bloch}.

Qubit errors can occur by a variety of physical processes. The simplest case to examine are errors which cause the qubit to coherently rotate from one point on the Bloch sphere to another. Such qubit errors could, for example, arise from systematic control faults in the hardware with which the qubits are realised. Mathematically, coherent errors are described by a unitary operation $U(\delta\theta,\delta\phi)$ which evolves the qubit state as follows
\begin{equation}\label{eq:coherent_noise}
U(\delta\theta,\delta\phi)\ket{\psi}=\cos{\frac{\theta +\delta \theta}{2}}\ket{0}+e^{i(\phi +\delta \phi)}\sin{\frac{\theta+\delta\theta}{2}}\ket{1}\rm,
\end{equation}    
where $\theta +\delta \theta$ and $\phi +\delta \phi$ are the new coordinates on the Bloch sphere. From this, we see that qubits are susceptible to a continuum of coherent errors obtained by varying the parameters $\delta \theta $ and $\delta \phi$. It would therefore seem, at first glance, that quantum error correction protocols should have to be based on techniques from classical analogue computation for which the theory of error correction is not well developed. Luckily, however, it turns out that quantum errors can be digitised so that the ability to correct for a finite set of errors is sufficient to correct for any error \cite{Knill97}. To see how this is possible, we first note that coherent noise processes are described by matrices that can be expanded terms of a Pauli basis.\footnote{For a more detailed definition of the Pauli group, and details of the Pauli notation used in this review, see appendix \ref{app:pauli}.} For example, the Pauli basis for two-dimensional matrices is given by 
\begin{equation}
\openone=\left( \begin{matrix}
1 & 0 \\ 0 & 1
\end{matrix} \right), \quad X=\left( \begin{matrix}
0 & 1 \\ 1 & 0
\end{matrix} \right), \quad Y=\left( \begin{matrix}
0 & -{\rm i} \\ {\rm i}& 0
\end{matrix} \right),  \quad Z=\left( \begin{matrix}
1 & 0 \\ 0 & -1
\end{matrix} \right) \rm.
\end{equation}
The single-qubit coherent error process described in equation (\ref{eq:coherent_noise}) can be expanded in the above basis as follows
\begin{equation}
U(\delta\theta,\delta\phi)\ket{\psi}=\alpha_{I}\openone \ket{\psi} + \alpha_{X} X \ket{\psi} + \alpha_ZZ \ket{\psi} + \alpha_Y Y \ket{\psi}
\end{equation}
where $\alpha_{I,X,Y,Z}$ are the expansion coefficients. By noting that the Pauli $Y$-matrix is equivalent (up to a phase) to the product $XZ$, this expression can be further simplified to
\begin{equation} \label{eq:dig_err}
U(\delta\theta,\delta\phi)\ket{\psi}=\alpha_{I}\openone \ket{\psi} + \alpha_{X} X \ket{\psi} + \alpha_ZZ \ket{\psi} + \alpha_{XZ} XZ \ket{\psi}\rm.
\end{equation}
The above expression shows that any coherent error process can be decomposed into a sum from the Pauli set $\{\openone, X, Z, XZ\}$. In the following sections, we will see that the error correction process itself involves performing projective measurements that cause the above superposition to collapse to a subset of its terms. As a result, a quantum error correction code with the ability to correct errors described the by the $X$- and $Z$-Pauli matrices will be able to correct any coherent error. This effect, referred to as the digitisation of the error, is crucial to the success of quantum error correction codes.

\subsection{Quantum error-types} \label{sec:errors}

As a result of the digitisation of the error there are two fundamental quantum error-types that need to be accounted for by quantum codes. Pauli $X$-type errors can be thought of as quantum bit-flips that map $X\ket{0}= \ket{1}$ and $X\ket{1}=\ket{0}$. The action of an $X$-error on the general qubit state is
\begin{equation}\label{eq:bit_flip_gen}
X\ket{\psi}=\alpha X\ket{0}+\beta X\ket{1}=\alpha\ket{1}+\beta\ket{0}\rm.
\end{equation}
The second quantum error type, the $Z$-error, is often referred to as a phase-flip and has no classical analogue. Phase-flips map the qubit basis states $Z\ket{0}=\ket{0}$ and $Z\ket{1}=-\ket{1}$, and therefore have the following action on the general qubit state
\begin{equation}\label{eq:phase_flip_gen}
Z\ket{\psi}=\alpha Z\ket{0}+\beta Z\ket{1}=\alpha\ket{0}-\beta\ket{1}\rm.
\end{equation}
So far, for simplicity, I have restricted discussion to coherent errors acting on single-qubits. However, the digitisation of the error result generalises to arbitrary quantum error processes, including those that describe incoherent evolution of the quantum state as a result of the qubits' interaction with their environment \cite{Knill97}.

\subsection{The challenges of quantum error correction}

The digitisation of quantum errors means it is possible to reuse certain techniques from classical coding theory in quantum error correction. However, there remain a number of complications that prevent the straight-forward translation of classical codes to quantum codes. The first complication is the no-cloning theorem for quantum states  \cite{Wootters82}, which asserts that it is not possible to construct a unitary operator $U_{\rm clone}$ which performs the following operation
\begin{equation} \label{eq:no_cloning}
U_{\rm clone}(\ket{\psi}\otimes \ket{0})\rightarrow \ket{\psi}\otimes\ket{\psi}\rm,
\end{equation}
where $\ket{\psi}$ is the state to be cloned. In contrast, classical codes work under the assumption that data can be arbitrarily duplicated. For quantum coding, it is therefore necessary to find alternative ways of adding redundancy to the system.

The second complication in quantum coding arises from the fact that qubits are susceptible to both bit-flips ($X$-errors) and phase-flips ($Z$-errors). Quantum error correction codes must therefore be designed with the ability to detect both error-types simultaneously. In contrast, in classical coding, only bit-flip errors need to be considered.

The final complication specific to quantum error correction is the problem of wavefunction collapse. In a classical system, it is possible to measure arbitrary properties of the bit register without risk of compromising the encoded information. For quantum codes, however, any measurements of the qubits performed as part of the error correction procedure must be carefully chosen so as not to cause the wavefunction to collapse and erase the encoded information. In the next section, we will see how this is achieved through the use of a special type of projective measurement referred to as a stabilizer measurement \cite{Gottesman97}.

\section{Quantum redundancy \& stabilizer measurement} \label{sec:2qubit}

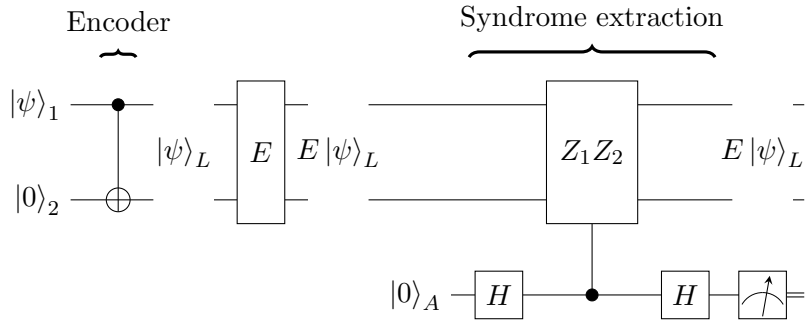
\begin{figure}
	\centering
	\input{figs/qec/two_qubit_v2.tikz}
	\caption{Circuit diagram for the two qubit code. Encode stage: the information contained in $\ket{\psi}_1$ is entangled with a redundancy qubit $\ket{0}_2$ to create a logical state $\ket{\psi}_L$. Error stage: during the error window (shown by the circuit element $E$), the two code qubits are potentially subject to bit-flip errors. Syndrome extraction stage: the $Z_1Z_2$ operator, controlled by the ancilla qubit $A$, is applied to the code qubits. The subsequent measurement of the ancilla gives the code syndrome $S$.}
	\label{fig:two_qubit}
\end{figure}

As outlined in the previous section, quantum error correction is complicated by the no-cloning theorem, wavefunction collapse and the existence of a uniquely quantum error-type, the phase-flip. So, faced with these challenges, how is redundancy added to a quantum system to allow errors to be detected in real time? Classical repetition codes work by increasing the resources used to encode the data beyond the theoretical minimum. Analogously, in quantum codes redundancy is added by expanding the Hilbert space in which the qubits are encoded \cite{Shor95}. To see how this is achieved in practice, we now describe the two-qubit code, a prototypical quantum code designed to detect a single-bit flip error. The encode stage of the two-qubit code, acting on the general state $\ket{\psi}$, has the following action
\begin{equation}
\ket{\psi}=\alpha\ket{0}+\beta\ket{1}\xrightarrow{\text{two-qubit encoder}}\ket{\psi}_L=\alpha \ket{00}+\beta\ket{11}=\alpha\ket{0}_L+\beta\ket{1}_L\rm,
\end{equation}
where after encoding the logical codewords are $\ket{0}_L=\ket{00}$ and $\ket{1}_L=\ket{11}$. Note that this does not correspond to cloning the state as
\begin{equation}
\ket{\psi}_L=\alpha \ket{00} + \beta \ket{11} \neq \ket{\psi}\otimes \ket{\psi}\rm.
\end{equation}
The effect of the encoding operation is to distribute the quantum information in the initial state $\ket{\psi}$ across the entangled two-party logical state $\ket{\psi}_L$. This introduces redundancy to the encoding that can be exploited for error detection. To understand exactly how this works, it is instructive to consider the computational Hilbert spaces before and after encoding. Prior to encoding, the single qubit is parametrised within a two-dimensional Hilbert space $\ket{\psi}\in \mathcal{H}_2=\text{span}\{\ket{0},\ket{1}\}$. After encoding the logical qubit occupies a four-dimensional Hilbert space
\begin{equation}
\ket{\psi}\in \mathcal{H}_4 =\text{span}\{\ket{00},\ket{01},\ket{10},\ket{11}\}\rm.
\end{equation}
More specifically the logical qubit is defined within a two-dimensional subspace of this expanded Hilbert space
\begin{equation}
\ket{\psi}_L\in\mathcal{C} = {\rm span}\{\ket{00},\ket{11}\} \subset \mathcal{H}_4\rm,
\end{equation}
where $\mathcal{C}$ is called the codespace. Now, imagine that the logical qubit is subject a bit-flip error on the first qubit resulting in the state
\begin{equation}
X_1 \ket{\psi}_L=\alpha \ket{10} + \beta \ket{01}\rm,
\end{equation}
where $X_1$ is a bit-flip error acting on the first qubit. The resultant state is rotated into a new subspace
\begin{equation}
X_1\ket{\psi}_L \in \mathcal{F} \subset \mathcal{H}_4\rm,
\end{equation}
where we call $\mathcal{F}$ the error subspace. Notice that an $X_2$-error will also rotate the logical state into the $\mathcal{F}$ subspace. If the logical state $\ket{\psi}_L$ is uncorrupted, it occupies the codespace $\mathcal{C}$, whereas if it has been subject to a single-qubit bit-flip, it occupies the error space $\mathcal{F}$. As the $\mathcal{C}$ and $\mathcal{F}$ subspaces are mutually orthogonal, it is possible to distinguish which subspace the logical qubit occupies via a projective measurement without compromising the encoded quantum information. In the context of quantum coding, measurements of this type are called stabilizer measurements.

For the purposes of differentiating between the codespace $\mathcal{C}$ and the error space $\mathcal{F}$, a projective measurement of the form $Z_1Z_2$ is performed. The $Z_1Z_2$ operator yields a $(+1)$ eigenvalue when applied to the logical state
\begin{equation}
Z_1 Z_2 \ket{\psi}_L = Z_1Z_2(\alpha \ket{00}+\beta \ket{11})=(+1)\ket{\psi}_L\rm.
\end{equation}
The $Z_1Z_2$ operator is said to \textit{stabilize} the logical qubit $\ket{\psi}_L$ as it leaves it unchanged \cite{Gottesman09}. Conversely, the $Z_1Z_2$ operator projects the errored states, $X_1\ket{\psi}_L$ and $X_2\ket{\psi}_L$, onto the $(-1)$ eigenspace. Notice that for either outcome, the information encoded in the $\alpha$ and $\beta$ coefficients of the logical state remains undisturbed.

\begin{table}
		\def\arraystretch{1.3}
	\tbl{The syndrome table for the two-qubit code. The syndrome $S$ is bit a string where each bit corresponds to the outcome of a stabilizer measurement.}
	{\begin{tabular}{lcc} \toprule
		
		Error & Syndrome, $S$ \\\midrule
		$I_1 I_2$ & $0$\\
		$X_1I_2$ & $1$\\
		$I_1 X_2$ & $1$\\
		$X_1 X_2$ & $0$\\
		
		  \bottomrule
	\end{tabular}}
	\label{tab:two_qubit}
\end{table}

Figure \ref{fig:two_qubit} shows the circuit implementation of the two-qubit code. In the encode stage, a \cnot gate is used to entangle the $\ket{\psi}$ state with a redundancy qubit to create the logical state $\ket{\psi}_L$. Following this, we assume the logical qubit is subject to a bit-flip error $E$, applied during the stage of the circuit labelled `E'. Following the error stage, an ancilla qubit $\ket{0}_A$ is introduced to perform the measurement of the $Z_1Z_2$ stabilizer. The syndrome extraction stage of the circuit transforms the quantum state as follows
\begin{equation} \label{eq:synd_extract}
E\ket{\psi}_L \ket{0}_A \xrightarrow[]{\text{syndrome extraction}} \dfrac{1}{2}(\openone_1\openone_2 + Z_1Z_2)E\ket{\psi}_L \ket{0}_A + \dfrac{1}{2}(\openone_1\openone_2 - Z_1Z_2)E\ket{\psi}_L \ket{1}_A\rm,
\end{equation}
where $E$ is an error from the set $\{\openone, X_1, X_2, X_1X_2\}$. Now, consider the case where $E=X_1$ so that the logical state occupies the error space $E\ket{\psi}_L \in \mathcal{F}$. In this scenario, it can be seen that the first term in equation (\ref{eq:synd_extract}) goes to zero. The ancilla qubit is therefore measured deterministically as `1'. Considering the other error patterns, we see that if the logical state is in the codespace (i.e., if $E=\{\openone, X_1X_2\}$) then the ancilla is measured as `0'. Likewise, if the logical state is in the error subspace (i.e., if $E=\{X_1,X_2\}$) then the ancilla is measured as `1'. The outcome of the ancilla qubit measurement is referred to as a \textit{syndrome}, and tells us whether or not the logical state has been subject to an error. The syndromes for all bit-flip error types in the two-qubit code are shown in table \ref{tab:two_qubit}.

Up to this point, we have assumed that the error introduced by the circuit element labelled `$E$' is deterministic. We now demonstrate how the two qubit code works under a more general probabilistic error of the type discussed in section \ref{sec:digitisation}. For the purposes of this example, we will assume that each qubit in the two-qubit code is subject to a coherent error of the form 
\begin{equation}
\mathcal{E}=\alpha_{I}\openone+\alpha_X X\rm,
\end{equation}
where $|\alpha_{I}|^2+|\alpha_X|^2=1$. Here we see that $|\alpha_X|^2=p_X$ is the probability of an $X$-error occurring on the qubit. The probability of no-error occurring is therefore equal to $|\alpha_I|^2=1-p_X$. The combined action of the error operator $\mathcal{E}$ acting on both qubits is given by
\begin{equation}
E=\mathcal{E}_1\otimes\mathcal{E}_2=\alpha_I^2\openone_1 \openone_2 + \alpha_I\alpha_X(X_1+X_2)+\alpha_X^2 X_1X_2\rm.
\end{equation}
With the above error operator $E$, the syndrome extraction stage in figure \ref{fig:two_qubit} stage transforms the quantum state as follows
\begin{equation} \label{eq:synd_extract_prob}
E\ket{\psi}_L \ket{0}_A \xrightarrow[]{\text{syndrome extraction}} (\alpha_I^2\openone_1 \openone_2 +\alpha_X^2 X_1X_2)\ket{\psi}_L\ket{0}_A + \alpha_I\alpha_X(X_1+X_2)\ket{\psi}_L\ket{1}_A\rm.
\end{equation}
If the syndrome is measured as `$0$', the state collapses to a subset of its terms 
\begin{equation} \label{eq:synd_extract_prob_collapse}
 \dfrac{(\alpha_I^2\openone_1 \openone_2 +\alpha_X^2 X_1X_2)}{\sqrt{|\alpha_I^2|^2+|\alpha_X^2|^2}}\ket{\psi}_L\ket{0}_A\rm,
\end{equation}
where the the denominator ensures normalisation. By calculating the square-norm in the first term in the above, we can calculate the probability $p_L$ that the logical state is subject to an error
\begin{equation}
p_L=\left|\dfrac{\alpha_X^2}{\sqrt{|\alpha_I^2|^2+|\alpha_X^2|^2}}\right|^2=\dfrac{p_x^2}{(1-p_x)^2+p_x^2}\approx p_x^2
\end{equation}
where the above approximation is made under the assumption that $p_x$ is small. For the single qubit $\ket{\psi}$, the probability of error is $p_x$ when it is subject to the error operator $\mathcal{E}$. For the logical qubit $\ket{\psi}_L$ subject to the error operator $\mathcal{E}_1\otimes\mathcal{E}_2$, the logical error rate is $p_L=p_x^2$. From this, we see that the two-qubit code suppresses the error rate relative to the un-encoded case.

\subsection{The three-qubit error correction code} \label{sec:three_qubit}

The syndrome produced by the two-qubit code informs us of the presence of an error, but does not provide enough information to allow us to infer which qubit the error occurred on. It is therefore a detection code. In order to create an error correction code with the ability to both detect and localise errors, multiple stabilizer measurements need to be performed.

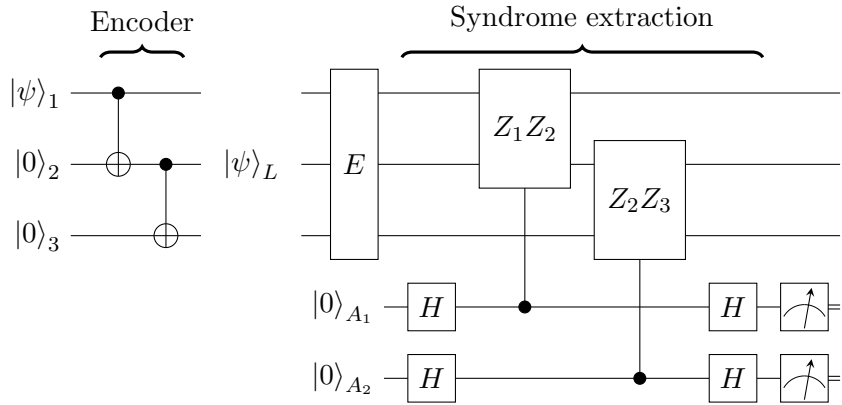
\begin{figure}
	\centering
	\input{figs/qec/three_qubit.tikz}
	\caption{The circuit diagram of the three-qubit code. Encode stage: The information contained in a single qubit $\ket{\psi}$ is entangled with two redundancy qubits $\ket{0}_2$ and $\ket{0}_3$ to create a logical qubit $\ket{\psi}_L$. The stabilizers $Z_1Z_2$ and $Z_2Z_3$ are measured on the logical qubit via two operations controlled on the ancilla qubits $A_1$ and $A_2$ respectively. The subsequent measurement of the ancilla qubits gives a two-bit syndrome $S$.}
	\label{fig:three_qubit_code}
\end{figure}

We now describe the three-qubit code, the natural extension of the two-qubit code in which the encoding operation distributes the quantum information across an entangled three-party state to give a logical state of the form $\ket{\psi}_L=\alpha\ket{000}+\beta\ket{111}$. This logical state occupies an eight-dimensional Hilbert space that can be partitioned into four two-dimensional subspaces as follows
\begin{equation}\begin{split}
&\mathcal{C}={\rm span}\{\ket{000},\ket{111}\}, \quad \mathcal{F}_1={\rm span}\{\ket{100},\ket{110}\}, \\ &\mathcal{F}_2={\rm span}\{\ket{010},\ket{101}\}, \quad \mathcal{F}_3 ={\rm span}\{\ket{001},\ket{110}\}\rm,
\end{split}
\end{equation}
where $\mathcal{C}$ is the logical code space, and $\mathcal{F}_{\{1,2,3\}}$ are the logical error spaces. We see that each single-qubit error from the set $E=\{X_1,X_2,X_3\}$ will rotate the codespace to a unique error space so that $X_i \ket{\psi}_L \in \mathcal{F}_i$. In order to differentiate between these subspaces, we perform two stabilizer measurements $Z_1Z_2$ and $Z_2Z_3$ via the circuit shown in figure \ref{fig:three_qubit_code}. The resultant syndrome table for single-qubit errors is given in table \ref{tab:three_qubit}. From this we see that each single-qubit error produces a unique two-bit syndrome $S=s_1s_2$, enabling us to choose a suitable recovery operation.

\subsection{Quantum code distance}

\begin{table}
	\def\arraystretch{1.3}
	\tbl{The syndrome table for all bit-flip erros on the three qubit code. The syndrome $S$ is a two-bit string formed by concatenating the results of the two stabilizer measurements.}
	{\begin{tabular}{lc|ccc} \toprule
			Error & Syndrome, $S$ & Error & Syndrome, $S$\\\midrule
			$I_1I_2I_3$ & $00$ & $X_1X_2I_3$ & $01$\\
			$X_1I_2I_3$ & $10$ & $I_1X_2X_3$ & $10$\\
			$I_1 X_2I_3$ & $11$ & $X_1I_2X_3$ & $11$\\
			$I_1 I_2 X_3$ & $01$ & $X_1X_2X_3$ & $00$\\
			\bottomrule
	\end{tabular}}
	\label{tab:three_qubit}
\end{table}

As is the case for classical codes, the distance of a quantum code is defined as the minimum size error that will go undetected. Alternatively, this minimum size error can be viewed as a logical Pauli operator that transforms one codeword state to another. For the three-qubit code described in section \ref{sec:three_qubit}, we see that the logical Pauli-$X$ operator is given by $\bar{X}=X_1X_2X_3$, so that
\begin{equation}
\bar{X}\ket{0}_L=\ket{1}_L \ {\rm and} \ \bar{X}\ket{1}_L=\ket{0}_L\rm,
\end{equation}
where $\ket{0}_L=\ket{000}$ and $\ket{1}_L=\ket{111}$ are the logical codewords for the three-qubit code. If it were the case that qubits were only susceptible to $X$-errors, then the three-qubit code would have distance $d=3$. However, as qubits are also susceptible to phase-flip errors, it is also necessary to consider the logical Pauli-$Z$ operator $\bar{Z}$ when determining the code distance. To do this, it is useful to switch from the computational basis, $\{\ket{0},\ket{1}\}$, to the conjugate basis, $ \{\ket{+},\ket{-}\}$, where we define
\begin{equation}
\ket{+}=\dfrac{1}{\sqrt{2}}(\ket{0}+\ket{1}) \quad \text{and} \quad \ket{-}=\dfrac{1}{\sqrt{2}}(\ket{0}-\ket{1})\rm .
\end{equation}
A $Z$-error maps the conjugate basis states as follows $Z\ket{+}=\ket{-}$ and $Z\ket{-}=\ket{+}$. Now, encoding the conjugate basis states with the three-qubit code gives the logical states
\begin{equation}
\ket{+}_L=\dfrac{1}{\sqrt{2}}(\ket{000}+\ket{111}) \quad \text{and} \quad \ket{-}_L=\dfrac{1}{\sqrt{2}}(\ket{000}-\ket{111})\rm.
\end{equation}
A weight-one logical Pauli-Z operator $\bar{Z}=Z_1$ will transform $Z\ket{+}_L=\ket{-}_L$, meaning the code is unable to detect the presence of single-qubit $Z$-errors. As a result, the three-qubit code has a quantum distance $d=1$. In the next section, we outline the construction of general stabilizer codes capable of detecting both $X$- and $Z$-errors.

\section{Stabilizer codes} \label{sec:stab_codes}

\begin{figure}
	{\input{figs/qec/stab_codes.tikz} }
	\caption{Circuit illustrating the structure of an $[[n,k,d]] $ stabilizer code. A quantum data register $\ket{\psi}_D=\ket{\psi_{1}\psi_{2}...\psi_{k}}$ is entangled with redundancy qubits $\ket{0}_R=\ket{0_{1}0_{2}...0_{{n-k}}}$ via an encoding operation to create a logical qubit $\ket{\psi}_L$. After encoding, a sequence of $n-k$ stabilizer checks $P_i$ are performed on the register, and each result copied to an ancilla qubit $A_i$. The subsequent measurement of the ancilla qubits provides an $m$-bit syndrome.}
	\label{fig:stab_code}
\end{figure}
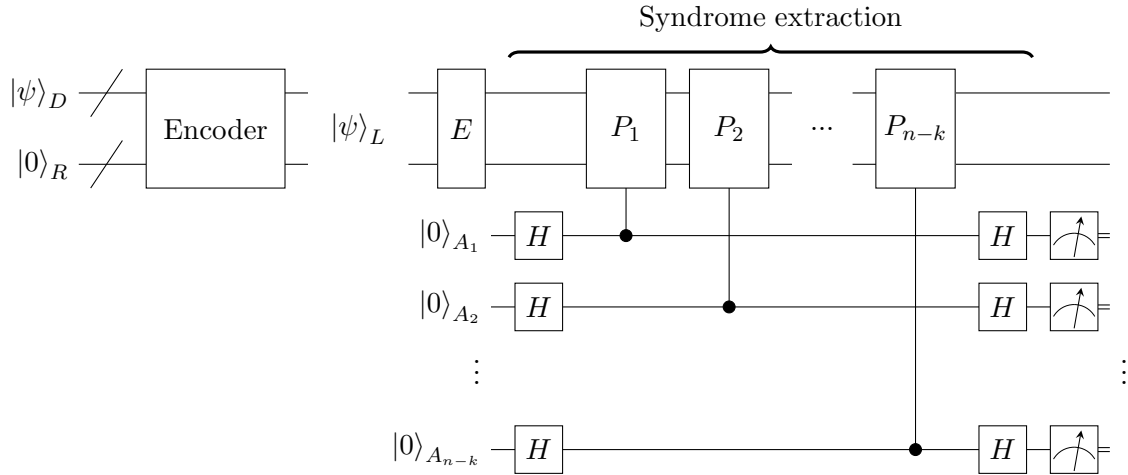

The three-qubit code works by de-localising the information in a single-qubit across three qubits. The resultant logical state is then encoded in a two-dimensional subspace (the codespace) of the expanded Hilbert space. The three-qubit code is designed such that if an $X$-error occurs, the logical state is rotated to an orthogonal error space, an event that can be detected via a sequence of two stabilizer measurements. This section describes how the procedure to can be generalised to create $[[n,k,d]]$ stabilizer codes, where $n$ is the total number of qubits, $k$ is the number of logical qubits and $d$ is the code distance. Note the use of double brackets to differentiate quantum codes from classical codes which are labelled with single brackets. 

The circuit in figure \ref{fig:stab_code} shows the basic structure of an $[[n,k,d]]$ stabilizer code. A register of $k$ data qubits, $\ket{\psi}_D$, is entangled with $m=n-k$ redundancy qubits $\ket{0}_R$ via an encoding operation to create a logical qubit $\ket{\psi}_L$. At this stage, the data previously stored solely in $\ket{\psi}_D$ is distributed across the expanded Hilbert space. Errors can then be detected by performing $m$ stabilizer measurements $P_i$ as shown to the right of figure \ref{fig:stab_code}.

In the circuit in figure \ref{fig:stab_code}, each of the stabilizers is measured using the same syndrome extraction method that was used for the two-qubit code in figure \ref{fig:two_qubit}. For each stabilizer $P_i$, the syndrome extraction circuit maps the logical state as follows
\begin{equation}\label{eq:synd_extract_gen}
E\ket{\psi}_L \ket{0}_{A_i} \xrightarrow[]{\text{syndrome extraction}} \dfrac{1}{2}(\openone^{\otimes n} + P_i)E\ket{\psi}_L \ket{0}_{A_i} + \dfrac{1}{2}(\openone^{\otimes n} - P_i)E\ket{\psi}_L \ket{1}_{A_i}\rm.
\end{equation}
From the above, we see that if the stabilizer $P_i$ commutes with an error $E$ the measurement of ancilla qubit $A_i$ returns `0'. If the stabilizer $P_i$ anti-commutes with an error $E$ the measurement returns `1'. The task of constructing a good code therefore involves finding stabilizers that anti-commute with the errors to be detected. In general, two Pauli operators will commute with one another if they intersect non-trivially on an even number of qubits, and anti-commute if otherwise. For specific examples of Pauli commutation relations, see appendix \ref{app:pauli_commute}.   

The results of the $m$ stabilizer measurements are combined to give an $m$-bit syndrome. For a well designed code, the syndrome allows us to deduce the best recovery operation to restore the logical state to the codespace.

\subsection{Properties of the code stabilizers}



The stabilizers $P_i$ of an $[[n,k,d]]$ code must satisfy the following properties:
\begin{enumerate}
\item They must be Pauli-group elements, $P_i \in \mathcal{G}_n$. Here $\mathcal{G}_n$ is the Pauli Group over $n$-qubits (see appendix \ref{app:pauli} for the definition of the Pauli Group).

\item They must stabilize all logical states $\ket{\psi}_L$ of the code. This means that each $P_i$ has the action $P_i\ket{\psi}_L=(+1)$ for all possible values of $\ket{\psi}_L$.

\item All the stabilizers of a code must commute with one another, so that $[P_i,P_j]=0$ for all $i$ and $j$. This property is necessary so that the stabilizers can be measured simultaneously (or in a way independent of their ordering) as depicted in figure \ref{fig:stab_code}.
  
\end{enumerate}
In the language of group theory, the stabilizers $P_i$ of an $[[n,k,d]]$ code form an Abelian subgroup $\mathcal{S}$ of the Pauli Group. The stabilizer requirements listed above are incorporated into the definition of $\mathcal{S}$ as follows
\begin{equation}\label{eq:stab_defs}
\mathcal{S}=\{P_i \in \mathcal{G}_n \ | \ P_i \ket{\psi}_L = (+1)\ket{\psi}_L \ \forall \ \ket{\psi}_L \wedge \ [P_i,P_j]=0 \ \forall \ (i,j)  \}\rm .
\end{equation}

An important point to note is that any product of the stabilizers $P_iP_j$ will also be a stabilizer as $P_iP_j\ket{\psi}_L=P_i(+1)\ket{\psi}_L=(+1)\ket{\psi}_L$. Given this, it is import to ensure that the set of $m=n-k$ stabilizers that are actually measured in the syndrome extraction process form a minimal set of the stabilizer group
\begin{equation}\label{eq:stab_generators}
\mathcal{S}=\langle G_1, G_2,...,G_m \rangle
\end{equation}
In a minimal set it is not possible to obtain one stabilizer $G_i$ as a product of any of the other elements $G_j$. As a simple example, consider the following set of stabilizers for the three-qubit code $\mathcal{S}=\{Z_1Z_2,Z_2Z_3,Z_1Z_3\}$. This is not a minimal set, as it is possible to obtain the third stabilizer as a product of the first two. A possible minimal set is $S=\langle Z_1Z_2,Z_2Z_3 \rangle$, which are the two stabilizers measured in the example in section \ref{sec:three_qubit}.


\subsection{The logical operators of stabilizer codes}\label{sec:logical_operators}

An $[[n,k,d]]$ stabilizer code has $2k$ logical Pauli operators that allow for logical states to be modified without having to decode then re-encode. For each logical qubit $i$, there is a logical Pauli-$X$ operator $\bar{X}_i$ and a logical Pauli-$Z$ operator $\bar{Z}_i$. Each pair of logical operators, $\bar{X}_i$ and $\bar{Z}_i$, satisfy the following properties
\begin{enumerate}
\item They commute with all the code stabilizers in $\mathcal{S}$.
\item They anti-commute with one another, so that $[\bar{X}_i,\bar{Z}_i]_{+}=\bar{X}_i\bar{Z}_i + \bar{Z}_i\bar{X}_i=0$ for all qubits $i$.
\end{enumerate}

Any product of a logical operator $\bar{L}_i$ and stabilizer $P_j$ will also be a logical operator. This is clear from the fact that the stabilizer maps the logical state onto its $(+1)$ eigenspace. Any product $\bar{L_i}P_j$ therefore has the following action on the logical state $\bar{L}_iP_j\ket{\psi}_L=\bar{L}_i\ket{\psi}_L$.

\subsection{Example: The [[4,2,2]] detection code}\label{sec:four_qubit}
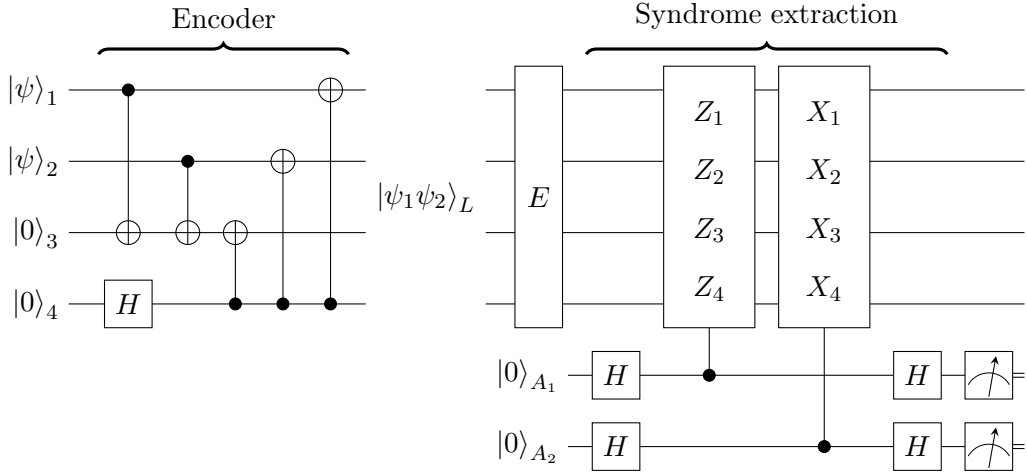
\begin{figure}
	\centering
	\input{figs/qec/422_code.tikz}
	\caption{Circuit diagram for the four-qubit code. Encode stage: the information contained in two-qubit register $\ket{\psi}_1\ket{\psi}_2$ is distributed across two redundancy qubits, $\ket{0}_3$ and $\ket{0}_4$, to create a logical state $\ket{\psi_1\psi_2}_L$ that encodes two qubits. Syndrome extraction stage: the code stabilizers, $Z_1Z_2Z_3Z_4$ and $X_1X_2X_3X_4$, are measured on the code qubits and results copied to the ancilla qubits. The subsequent measurement of the ancilla qubits provides a two-bit syndrome $S$ that informs of the occurrence of an error.}
	\label{fig:four_qubit}
\end{figure}

The $[[4,2,2]]$ detection code is the smallest stabilizer code to offer protection against a quantum noise model in which the qubits are susceptible to both $X$- and $Z$-errors \cite{Vaidman96,Grassl97}. As such, it provides a useful demonstration of the structure and properties of a stabilizer code.

An encoder for the $[[4,2,2]]$ code is shown in figure \ref{fig:four_qubit}. A two-qubit register $\ket{\psi}_1\ket{\psi}_2$ is entangled across four qubits to give the code state $\ket{\psi_1 \psi_2}_L$. As there are two encoded logical qubits in the $[[4,2,2]]$ code, its codespace is four-dimensional and is spanned by
\begin{equation}
\mathcal{C}_{[[4,2,2]]} = {\rm span}\left\{ \begin{matrix}
\ket{00}_L=\frac{1}{\sqrt{2}}(\ket{0000}+\ket{1111})\\
\ket{01}_L=\frac{1}{\sqrt{2}}(\ket{0110}+\ket{1001},\\
\ket{10}_L=\frac{1}{\sqrt{2}}(\ket{1010}+\ket{0101})\\
\ket{11}_L=\frac{1}{\sqrt{2}}(\ket{1100}+\ket{0011})
\end{matrix}
\right\}\rm.
\end{equation}
The stabilizers of the above logical basis states are $\mathcal{S}_{[[4,2,2]]} = \langle X_1X_2X_3X_4,Z_1Z_2Z_3Z_4 \rangle$. It is clear that these stabilizers commute with one another, as required by the definition in equation (\ref{eq:stab_defs}). These stabilizers can be measured using the syndrome extraction circuit shown to the right of figure \ref{fig:four_qubit}. From equation (\ref{eq:synd_extract_gen}), we know that for a syndrome measurement to be non-zero, the error $E$ has to anti-commute with the stabilizer being measured. Considering first the single-qubit $X$-errors ($E=\{X_1,X_2,X_3,X_4\}$), we see that they all anti-commute with the $Z_1Z_2Z_3Z_4$ stabilizer. Likewise, the single-qubit $Z$-errors $(E=\{Z_1,Z_2,Z_3,Z_4\}$) anti-commute with the $X_1X_2X_3X_4$ stabilizer. Any single-qubit error on the $[[4,2,2]]$ code will therefore trigger a non-zero syndrome. The syndrome table for all single-qubit errors in the $[[4,2,2]]$ code is shown in table \ref{tab:four_qubit}. For completeness, table \ref{tab:four_qubit} also includes the syndromes for single-qubit $Y$-errors, which are equivalent to the simultaneous occurrence of an $X$- and $Z$-error.

The $[[4,2,2]]$ code has Pauli-$X$ and Pauli-$Z$ logical operators for each of its encoded logical qubits. A possible choice of these logical operators is given by
\begin{equation}
\mathcal{L}_{[[4,2,2]]}=\left\{
\begin{matrix}
\bar{X}_1=X_1X_3\\
\bar{Z}_1=Z_1Z_4\\
\bar{X}_2=X_2X_3\\
\bar{Z}_2=Z_2Z_4
\end{matrix}
\right\}\rm.
\end{equation}
Each logical operators commutes with the two stabilizers of the code so that $\left[L_i, P_i\right]=0$ for all $L_i\in\mathcal{L}_{[[4,2,2]]}$ and $P_i\in\mathcal{S}$. Furthermore, it can be checked that the requirement $\left[ \bar{X}_i, \bar{Z}_i \right]_{+}=0$ is satisfied for each pair of logical operators. The minimum weight logical operator in $\mathcal{L}_{[[4,2,2]]}$ is two, which sets the code distance to $d=2$. As the distance of the $[[4,2,2]]$ code is less than three, it is a detection code rather than a full correction code. In section \ref{sec:shor9}, we introduce the Shor $[[9,1,3]]$ code as an example of a code capable of both detecting and correcting errors.

\begin{table}
	\def\arraystretch{1.3}
	\tbl{The syndrome table for the $[[4,2,2]]$ code for all single-qubit $X$-, $Z$- and $Y$-errors.}
	{\begin{tabular}{lc|cc|ccc} \toprule
			Error & Syndrome, $S$ & Error & Syndrome, $S$& Error & Syndrome, $S$\\\midrule
			$X_1$ & $10$ & $Z_1$ & $01$& $Y_1$ & $11$\\
			$X_2$ & $10$ & $Z_2$ & $01$& $Y_2$ & $11$\\
			$X_3$ & $10$ & $Z_3$ & $01$& $Y_3$ & $11$\\
			$X_4$ & $10$ & $Z_4$ & $01$& $Y_4$ & $11$\\
			\bottomrule
	\end{tabular}}
	\label{tab:four_qubit}
\end{table}

\subsection{A general encoding circuit for stabilizer codes}\label{sec:gen_encoder}

The quantum codes presented this review have included bespoke encoding circuits to prepare the logical states. Special methods exist for constructing such circuits given a set of stabilizers \cite{Gottesmanthesis,Roffe18}. In this section we describe a general method for preparing the logical states of stabilizer code using the same circuits that are used for syndrome extraction. 

The $\ket{0}_L$ codeword of any $[[n,k,d]]$ stabilizer can be obtained via a projection onto the $(+1)$ eigenspace of all of its stabilizers
\begin{equation}\label{eq:projection}
\ket{0}_L=\dfrac{1}{N}\prod_{P_i\in\langle \mathcal{S}\rangle}(\openone^{\otimes n} + P_i)\ket{0^{\otimes n}}\rm,
\end{equation}  
where $\langle \mathcal{S} \rangle$ is the minimal set of the code stabilizers and the $1/N$ term is a factor that ensures normalisation. For example, the $\ket{00}_L$ codeword of the four-qubit code defined in section \ref{sec:four_qubit} is given by
\begin{equation}
\ket{00}_L=\dfrac{1}{\sqrt{2}}(\openone^{\otimes 4} + X_1X_2X_3X_4)(\openone^{\otimes 4}+Z_1Z_2Z_3Z_4)\ket{0000}=\dfrac{1}{\sqrt{2}}(\ket{0000}+\ket{1111})\rm.
\end{equation}
The remaining codewords of the code can be obtained by applying logical operators to the $\ket{0}_L$ codeword.

The $\ket{0}_L$ codeword of any stabilizer code can be prepared via the projection in equation (\ref{eq:projection}) by applying the general syndrome extraction circuit (shown on the right-hand-side of figure \ref{fig:stab_code}) to a $\ket{0}^{\otimes n}$ state. As an example, consider the case where we apply the syndrome extraction circuit to the state $\ket{0}^{\otimes 4}$ to prepare the $\ket{0}_L$ codeword of the four-qubit code. The intermediary state immediately after the extraction of the $X_1X_2X_3X_4$ stabilizer is given by
\begin{equation}\label{eq:four_qubit_codeword}
\dfrac{1}{2}(\openone^{\otimes 4} + X_1X_2X_3X_4)\ket{0000}\ket{0}_A + \dfrac{1}{2}(\openone^{\otimes 4} - X_1X_2X_3X_4)\ket{0000}\ket{1}_A\rm.
\end{equation}
When the ancilla is measured, the above state collapses to either the $(+1)$ or $(-1)$ projection with equal probability. In the case where the `$1$' syndrome is measured, a correction needs to be applied to transform the state back onto the $(+1)$ eigenspace of the stabilizer. Repeating this procedure for the remaining stabilizers leads to the preparation of the $\ket{0}_L$ codeword.

\subsection{Quantum error correction with stabilizer codes} \label{sec:qec_proc}
As is the case for classical codes, the distance of a quantum code is related to the number of correctable errors $t$ via the relation $d=2t+1$. As a result, stabilizer codes with $d \geq 3$ are error correction codes for which active recovery operations can be applied. In contrast, detection protocols such as the $[[4,2,2]]$ code require a repeat-until-success approach.

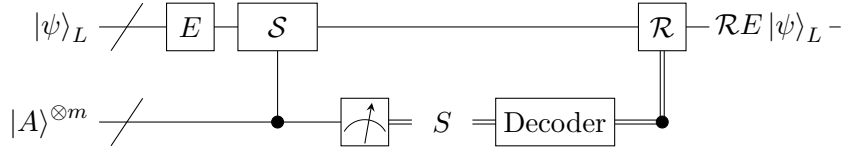
\begin{figure}
\centering
\input{figs/qec/qec_proc2.tikz}
\caption{The general procedure for active recovery in a quantum error correction code. The logical qubit $\ket{\psi}_L$ of an $[[n,k,d]]$ stabilizer code is subject to an error process $E$. A generating set of stabilizers $\mathcal{S}$ are measured on the logical state to yield an $m$-bit syndrome $S$. This syndrome is processed by a decoder to determine the best recovery operation $\mathcal{R}$ to return the logical state to the codespace. After the recovery has been applied, the output of the error correction cycle is $\mathcal{R}E\ket{\psi}_L$. Double lines indicate classical information flow.}
\label{fig:qec_proc}
\end{figure}

Figure \ref{fig:qec_proc} shows the general error correction procedure for a single cycle of an $[[n,k,d\geq3]]$ stabilizer code. The encoded logical state $\ket{\psi}_L$ is subject to an error process described by the circuit-element $E$. Next, the code stabilizers are measured (using the syndrome extraction method illustrated in figure \ref{fig:stab_code}), and the results copied to a register of $m=n-k$ ancilla qubits $\ket{A}^{\otimes m}$. The ancilla qubits are then read out to give an $m$-bit syndrome $S$.

The next step in the error correction procedure is referred to as decoding, and involves processing the syndrome to determine the best unitary operation $\mathcal{R}$ to return the logical state to the codespace. After this recovery operation has been applied, the output of the code-cycle is given by $
\mathcal{R}E\ket{\psi}_L \in \mathcal{C}_{[[n,k,d]]}$. The decoding step is a success if the combined action of $\mathcal{R}E$ on the code state is as follows 
\begin{equation}\label{eq:qec_success}
\mathcal{R}E\ket{\psi}_L=(+1)\ket{\psi}_L\rm.
\end{equation}
The above condition is trivially satisfied if $\mathcal{R}=E^\dagger$ so that $\mathcal{R}E=\openone$. However, this is not the only solution. Equation (\ref{eq:qec_success}) is also satisfied for any product $\mathcal{R}E$ that is an element of the code stabilizer such that $\mathcal{R}E=P\in \mathcal{S}$. In section \ref{sec:shor9}, we will see that the fact that the solution for $\mathcal{R}$ is not unique means it is possible to design \textit{degenerate} quantum codes for which multiple errors can map to the same syndrome.

The decoding step fails if the recovery operation maps the code state as follows
\begin{equation}
\mathcal{R}E\ket{\psi}_L=L\ket{\psi}_L\rm,
\end{equation}
where $L$ is a logical operator of the code. In this case, the state is returned to the codespace, but the recovery operation leads to a change in the encoded information.

\subsection{Example: The Shor [[9,1,3]] code}\label{sec:shor9}

The Shor nine-qubit code was the first quantum error correction scheme to be proposed. It is an example of a distance-three degenerate code for which it is possible to apply a successful recovery operation for any single-qubit error \cite{Shor95}. We now outline how the Shor code can be constructed via a method known as code concatenation.

Code concatenation involves embedding the output of one code into the input of another. In the construction of the Shor nine-qubit code, the two codes that are concatenated are the three-qubit code for bit-flips and the three-qubit code for phase-flips \cite{Terhal15}. The three-qubit code for bit-flips $\mathcal{C}_{\rm 3b}$ was described in section \ref{sec:three_qubit} and is defined as follows
\begin{equation}\begin{split}
\mathcal{C}_{\rm 3b}={\rm span}\{\ket{0}_{\rm 3b}=\ket{000}, \ket{1}_{\rm 3b}=\ket{111}\}, \quad \mathcal{S}_{\rm 3b}=\langle Z_1Z_2,Z_2Z_3 \rangle\rm,
\end{split}
\end{equation}
where $\mathcal{S}_{\rm 3b}$ are the code stabilizers. Similarly, the three-qubit code for phase-flips $\mathcal{C}_{\rm 3p}$ is defined 
\begin{equation}
\begin{split}
\mathcal{C}_{\rm 3p}={\rm span}\{\ket{0}_{\rm 3p}=\ket{+++}, \ket{1}_{\rm 3p}=\ket{---}\}, \quad \mathcal{S}_{\rm 3p}=\langle X_1X_2,X_2X_3\rangle\rm,
\end{split}
\end{equation}
To construct the nine-qubit code, the bit-flip code is embedded into the codewords of the phase-flip code. This concatenation maps the $\ket{0}_{\rm 3p}$ codeword of the phase-flip code to a nine-qubit codeword $\ket{0}_{9}$ as follows
\begin{equation} \label{eq:concat1}
\ket{0}_{\rm 3p}=\ket{+++}\xrightarrow{\rm concatenation} \ket{0}_9=\ket{+}_{\rm 3b}\ket{+}_{\rm 3b}\ket{+}_{\rm 3b}\rm,
\end{equation}
where $\ket{+}_{\rm 3b}=\dfrac{1}{\sqrt{2}}(\ket{000}+\ket{111})$ is a logical state of the bit-flip code. Similarly, the concatenation maps the $\ket{1}_{\rm 3p}$ codeword of the phase-flip code to
\begin{equation}\label{eq:concat2}
\ket{1}_{\rm 3p}=\ket{---}\xrightarrow{\rm concatenation} \ket{1}_9=\ket{-}_{\rm 3b}\ket{-}_{\rm 3b}\ket{-}_{\rm 3b}\rm,
\end{equation}
where $\ket{-}_{\rm 3b}=\dfrac{1}{\sqrt{2}}(\ket{000}-\ket{111})$. The code defined by the codewords $\ket{0}_9$ and $\ket{1}_9$ is the nine-qubit Shor code with paramaters $[[9,1,3]]$. Rewriting the right-hand-sides of Equations \ref{eq:concat1} and \ref{eq:concat2} in the computational basis, we get the following codespace for the Shor code  
\begin{equation}\mathcal{C}_{[[9,1,3]]}={\rm span}\left\{\begin{matrix}
\ket{0}_9=\dfrac{1}{\sqrt{8}}(\ket{000}+\ket{111})(\ket{000}+\ket{111})(\ket{000}+\ket{111})\\
\ket{1}_9=\dfrac{1}{\sqrt{8}}(\ket{000}-\ket{111})(\ket{000}-\ket{111})(\ket{000}-\ket{111})
\end{matrix}\right\}\rm.
\end{equation}
The stabilizers of the above code are given by
\begin{equation}\begin{split}
\mathcal{S}_{[[9,3,3]]}=\langle Z_1Z_2,Z_2Z_3,Z_4Z_5,Z_5Z_6,Z_7Z_8,Z_8Z_9,\\X_1X_2X_3X_4X_5X_6,X_4X_5X_6X_7X_8X_9\rangle\rm.
\end{split}
\end{equation}
The first six terms are the stabilizers of the bit-flip codes in the three-blocks of the code. The final two stabilizers derive from the stabilizers of the phase-flip code.

Table \ref{tab:9qubit} shows the syndromes for all single-qubit errors in the nine-qubit code. Each of the $X$-errors produce unique syndromes. In contrast, $Z$-errors that occur in the same block of the code have the same syndrome. Fortunately, this degeneracy in the code syndromes does not reduce the code distance. To see why this is the case, consider the single-qubit errors $Z_1$ and $Z_2$, both of which map to the syndrome `$00000010$'. The decoder therefore has insufficient information to differentiate between the two errors, and will output the same recovery operation for either. For the purposes of this example, we will assume that the recovery operation the decoder outputs is $\mathcal{R}=Z_1$. For the case where the error is $E=Z_1$, the recovery operation trivially restores the logical state as $\mathcal{R}E\ket{\psi}_9= Z_1Z_1\ket{\psi}_9=\ket{\psi}_9$. In the event where $E=Z_2$, the recovery operation still restores the logical state as $\mathcal{R}E=Z_1Z_2$ is in the stabilizer of $\mathcal{C}_{[[9,1,3]]}$, and therefore acts on the logical state as follows $Z_1Z_2\ket{\psi}_9=\ket{\psi}_9$. The same arguments can be applied to the remaining degenerate errors of the code. As a result, the nine-qubit code has the ability to correct all single-qubit errors and has distance $d=3$.

\begin{table}
	\def\arraystretch{1.3}
	\tbl{The syndrome table for single-qubit $X$- and $Z$-errors on the nine-qubit code. The nine-qubit code is a degenerate code, as certain $Z$-errors share the same syndrome.}
	{\begin{tabular}{lc|ccc} \toprule
			
			Error & Syndrome, $S$ & Error & Syndrome, $S$\\\midrule
			$X_1$ & $10000000$ & $Z_1$ & $00000010$\\
			$X_2$ & $11000000$ & $Z_2$ & $00000010$\\
			$X_3$ & $01000000$ & $Z_3$ & $00000010$\\
			$X_4$ & $00100000$ & $Z_4$ & $00000011$\\
			$X_5$ & $00110000$ & $Z_5$ & $00000011$\\
			$X_6$ & $00010000$ & $Z_6$ & $00000011$\\
			$X_7$ & $00001000$ & $Z_7$ & $00000001$\\
			$X_8$ & $00001100$ & $Z_8$ & $00000001$\\
			$X_9$ & $00000100$ & $Z_9$ & $00000001$\\			
			\bottomrule
	\end{tabular}}
	\label{tab:9qubit}
\end{table}

\section{The surface code}\label{sec:surface}

The challenge in creating quantum error correction codes lies in finding commuting sets of stabilizers that enable errors to be detected without disturbing the encoded information. Finding such sets is non-trivial, and special code constructions are required to find stabilizers with the desired properties. In section \ref{sec:shor9} we saw how a code can be constructed by concatenating two smaller codes. Other constructions include methods for repurposing classical codes to obtain commuting stabilizer checks \cite{Calderbank95,Steane96b,Kovalev13,Tillich14}. In this section, we outline a construction known as the surface code \cite{Bravyi98,Freedman98}.

The realisation of a surface code logical qubit is key goal for many quantum computing hardware efforts \cite{Nickerson14,Kelly16,Rigetti16,OGorman16,Takita17}. Surface codes belong to a broader family of so-called \textit{topological} codes \cite{Kitaev03}. The general design principle behind topological codes is that the code is built up by `patching' together repeated elements. We will see that this modular approach ensures that the surface code can be straight-forwardly scaled in size whilst ensuring stabilizer commutativity. In terms of actual implementation, the specific advantage of surface code for current hardware platforms is that it requires only nearest-neighbour interactions. This is advantageous as many quantum computing platforms are unable to perform high-fidelity long-range interactions between qubits.

\subsection{The surface code four-cycle}

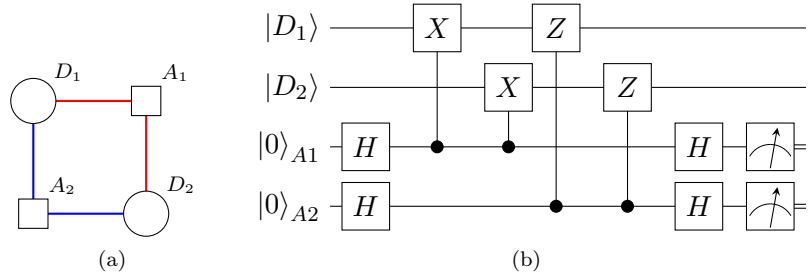
\begin{figure}
	\centering
	\subfloat[]{%
		\resizebox*{!}{!}{\input{tikz/surface_15.tikz}}}\hspace{5pt}
	\subfloat[]{%
		\resizebox*{!}{!}{\input{tikz/four_cycle.tikz}}}

	\caption{The surface code four-cycle. (a) Pictorial representation. The code qubits, $D_1$ and $D_2$, are represented by the circular nodes. The ancilla qubits, $A_1$ and $A_2$, are represented by the square nodes. The red and blue edges depict controlled-$X$ and controlled-$Z$ operations controlled on the ancilla qubits and acting on the code qubits. (b) An equivalent surface code four-cycle in circuit notation.}
	
	\label{fig:surface_15}
\end{figure}

For surface codes it is beneficial to adopt a pictorial representation of the code qubits in place of the circuit notation we have used up to this point. Figure \ref{fig:surface_15}a shows a surface code four-cycle, the fundamental building block around which surface codes are constructed. The circles in figure \ref{fig:surface_15}a represent the code qubits and the squares the ancilla qubits. The red edges represent controlled-$X$ gates, each controlled on an ancilla qubit $A$ and acting on a data qubit $D$. Likewise, the blue edges represent controlled-$Z$ operations, each controlled by an an ancilla qubit and acting on a data qubit. These controlled operations are the gates with which the stabilizers of the four-cycle are measured. Ancilla qubit $A_1$ connects to data qubits $D_1$ and $D_2$ via red edges, and therefore measures the stabilizer $X_{D_1}X_{D_2}$. Likewise, ancilla qubit $A_2$ measures the stabilizer $Z_{D_1}Z_{D_2}$. For comparison, the four-cycle is shown in quantum circuit notation in figure \ref{fig:surface_15}b.

The stabilizers of the four-cycle, $X_{D_1}X_{D_2}$ and $Z_{D_1}Z_{D_2}$, commute with one another as they intersect non-trivially on an even number of code qubits. This can easily be verified by inspection of figure \ref{fig:surface_15}b.

The $\ket{0}_L$ codeword of the four-cycle can be prepared by setting the initial state of the code qubits to $\ket{D_1D_2}=\ket{00}$, and following the general encoding procedure outlined in section \ref{sec:gen_encoder}. However, as the four-cycle has two code qubits $n=2$ and two stabilizers $m=2$, the number of logical qubits it encodes is equal to $k=n-m=0$. As a result, the four-cycle is not in itself a useful code. However, we will see that working detection and correction codes can be formed by tiling together multiple four-cycles to form square lattices.

\subsection{The $[[5,1,2]]$ surface code}

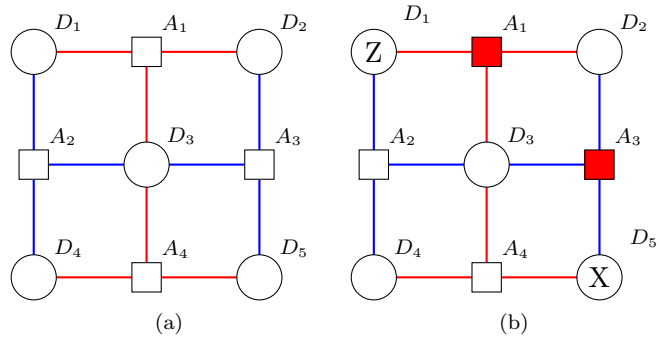
\begin{figure}
	\centering
	\subfloat[]{%
		\resizebox*{!}{!}{\input{tikz/surface_2.tikz}}}\hspace{5pt}
	\subfloat[]{%
		\resizebox*{!}{!}{\input{tikz/surface_2_error.tikz}}}

	\caption{(a) The $[[5,1,2]]$  surface code formed by tiling together four four-cycles in a square lattice. (b) Examples of error detection in the $[[5,1,2]]$ surface code. The $Z_{D_1}$ error on qubit $D_1$ anti-commutes with the stabilizer measured by ancilla qubit $A_1$. The $A_1$ qubit is coloured red to indicate it will measured as a `1'. Likewise, the $X_{D_5}$ error on qubit $D_5$ is detected by the stabilizer measured by ancilla qubit $A_3$.  }
	
	\label{fig:surface_2}
\end{figure}

\begin{figure}
	\centering
	\subfloat[]{%
		\resizebox*{!}{!}{\input{tikz/surface_2_lx.tikz}}}\hspace{5pt}
	\subfloat[]{%
		\resizebox*{!}{!}{\input{tikz/surface_2_lz.tikz}}}

	\caption{The logical operators of a surface code can be defined as chains of Pauli operations that act along the boundaries of the lattice. (a) The Pauli-$X$ logical operator $\bar{X}=X_{D_1}X_{D_4}$ acts along the boundary along which $Z$-type stabilizers are measured. (b) The Pauli-$Z$ logical operator $\bar{Z}=Z_{D_1}Z_{D_2}$ acts along the boundary along which $X$-type stabilizers are measured. The two logical operators anti-commute with one another.}
	
	\label{fig:surface_2_logicals}
\end{figure}
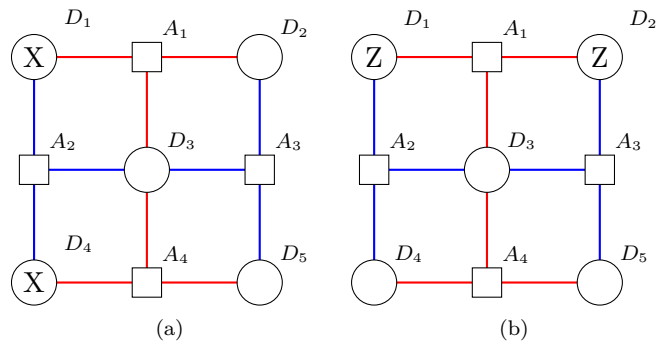

Figure \ref{fig:surface_2}a shows the five-qubit surface code formed by tiling together four four-cycles in a square lattice \cite{Horsman12}. By inspecting which data qubits each ancilla qubit connects to, the stabilizers of the code in figure \ref{fig:surface_2} can be read off to give
\begin{equation}\label{eq:512stabs}
\mathcal{S}_{\rm [[5,1,2]]} =\langle X_{D_1}X_{D_2}X_{D_3}, \ Z_{D_1}Z_{D_3}Z_{D_4}, \ Z_{D_2}Z_{D_3}Z_{D_5}, \ X_{D_3}X_{D_4}X_{D_5}\rangle \rm .
\end{equation}
The first term in the above is the stabilizer measured by ancilla qubit $A_1$, the second by ancilla $A_2$ etc. The stabilizers in $\mathcal{S}_{\rm [[5,1,2]]}$ commute with one another, as the $X$- and $Z$-type stabilizers all intersect on an even number of code qubits. From figure \ref{fig:surface_2}, we see that there are five code qubits and four stabilizers meaning the code encodes one logical qubit.

Figure \ref{fig:surface_15}b shows two examples of errors on the surface code and how they are detected. The $Z_{D_1}$-error on qubit $D_1$ anti-commutes with the $X_{D_1}X_{D_2}X_{D_3}$ stabilizer, and therefore triggers a `1' syndrome. This is depicted by the red filling in the ancilla qubit $A_1$. Likewise, the $X_{D_5}$-error anti-commutes with the $Z_{D_2}Z_{D_3}Z_{D_5}$ stabilizer and triggers a `1' syndrome measurement in ancilla qubit $A_4$.

From figure \ref{fig:surface_15} it can be seen that the surface code is a square lattice with two types of boundaries. The vertical boundaries are formed of blue edges representing $Z$-type stabilizer measurements. The horizontal boundaries are formed of red-edges representing $X$-type stabilizer measurements. The logical operators of the surface code can be defined as chains of Pauli operators along the edges of these boundaries.

Figure \ref{fig:surface_2_logicals}a shows a two-qubit Pauli chain $X_{D_1}X_{D_4}$ along the left-hand boundary of the five-qubit surface code. The $X_{D_1}X_{D_4}$ operator commutes with all the stabilizers in $\mathcal{S}_{[[5,1,2]]}$, in particular the stabilizer $Z_{D_1}Z_{D_3}Z_{D_4}$ with which is shares two qubits. Similarly, figure \ref{fig:surface_2_logicals}b shows an operator $Z_{D_1}Z_{D_2}$ which acts across the top of the lattice. It can easily be checked that this operator also commutes with all the code stabilizers. Finally, we note that the operators $X_{D_1}X_{D_4}$ and $Z_{D_1}Z_{D_2}$ anti-commute. As outlined in section \ref{sec:logical_operators}, the Pauli-$X$ and Pauli-$Z$ logical operators for each encoded qubit are pairs of operators that commute with all the code stabilizers but anti-commute with one another. A suitable choice for the logical operators of the $[[5,1,2]]$ surface code would therefore be
\begin{equation}
\bar{X}=X_{D_1}X_{D_4} \ {\rm and} \ \bar{Z}=Z_{D_1}Z_{D_2}\rm.
\end{equation}
From the above we see that the minimum weight of the logical operators is $2$, meaning the $[[5,1,2]]$ code is a detection code with $d=2$.

\subsection{Scaling the surface code}

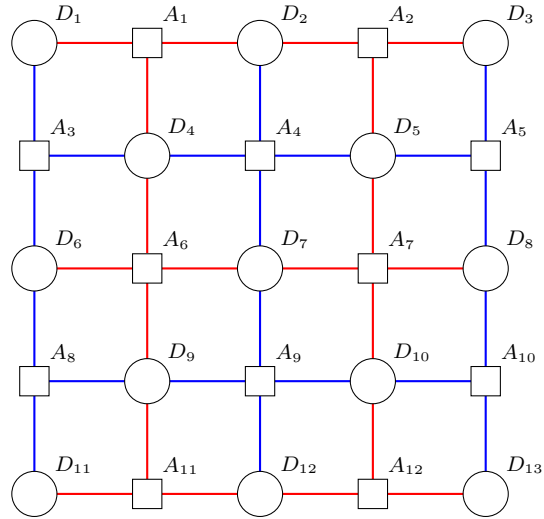
\begin{figure}
\centering
\input{tikz/surface_3.tikz}
\caption{A distance-three surface code with parameters $[[13,1,3]]$. A possible choice for the logical operators of this code would be $\bar{X}=X_{D_1}X_{D_6}X_{D_{11}}$ and $\bar{Z}=Z_{D_1}Z_{D_2}Z_{D_3}$.}
\label{fig:surface_3}
\end{figure}

The distance of a surface code can be increased simply by scaling the size of lattice. In general, a surface code with distance $d=\lambda$ will encode a \textit{single} logical qubit and have code parameters given by 
\begin{equation}
[[n=\lambda^2+(\lambda -1)^2,k=1,d=\lambda]]\rm .
\end{equation}
For example, the distance-three $[[13,1,3]]$ surface code is depicted in figure \ref{fig:surface_3}. The Pauli-$X$ logical operator of a surface code can be defined as a chain of $X$-Pauli operators along the boundary of the code along which the $Z$-stabilizers are applied (the blue boundary in our pictorial representation). Likewise, the $Z$-Pauli logical operator can be defined as a chain of $Z$-operators across the adjacent boundary along which the $X$-type stabilizers are applied (the red edges in our pictorial representation). For the distance-three code, a choice of logical operators would be $\bar{X}=X_{D_1}X_{D_6}X_{D_{11}}$ and $\bar{Z}=Z_{D_1}Z_{D_2}Z_{D_3}$. The $[[13,1,3]]$ code is the smallest surface code capable of detecting and correcting errors.

\section{Practical considerations for quantum error correction}\label{sec:practical}

Up to this point, we have described stabilizer codes in an idealised, theoretical setting. In this section, we outline some of the practical issues that arise when considering the implementation of quantum error correction codes on actual hardware.

\subsection{Efficient decoding algorithms}\label{sec:decode}

Given a code syndrome $S$, the role of the decoder is to find the best recovery operation $\mathcal{R}$ to restore the encoded quantum information to the codespace. Measuring the stabilizers of an $[[n,k,d]]$ code will produce an $m$-bit syndrome where $m=n-k$. As a result, there are $2^m$ possible syndromes for each code. For the small code examples described in this review, it is possible to compute lookup-tables  that exhaustively list the best recovery operation for each of the $2^m$ syndromes. However, such a decoding strategy rapidly becomes impractical as the code size is increased. As an example, consider the distance-five surface code which has parameters $[[41,1,5]]$. This code produces syndromes of length $m=40$, and would therefore need a lookup table of size $2^{40}\approx 10^{12}$.

In place of lookup tables, large-scale quantum error correction codes use approximate inference techniques to determine the most likely error to have occurred given a certain syndrome $S$. Such methods allow for recovery operations to be chosen and applied in real-time between successive stabilizer code cycles. Unfortunately, there is no known universal decoder that can be efficiently applied to all quantum error correction codes. Instead, bespoke decoding algorithms need to be developed that are designed for specific code constructions. For surface codes, a technique known as minimum weight perfect matching (MWPM) can be used for decoding, which works by identifying error chains between positive syndrome measurements \cite{Edmonds65,Kolmogorov09}.

As outlined in section \ref{sec:qec_proc}, the decode stage of an error correction cycle fails when $\mathcal{R}E=L$, where $\mathcal{R}$ is the recovery operation output by the decoder, $E$ is the error and $L$ is a logical operator of the code. The frequency with which the decoder fails in this way gives a logical error rate $p_L$. As decoding algorithms are based on approximate inference techniques, some perform better than others. As such, the logical error rate of a quantum error correction code will depend heavily on the decoder used. The logical error rate can be determined by simulating stabilizer code cycles with errors sampled from a noise model. The specifics of the noise model are motivated by the physical device on which the code is to be run.

\subsection{Code thresholds}
A code construction provides a method for building a set of codes with a shared underlying structure. An example are the surface codes, for which the code distance can be increased by expanding the size of the qubit lattice. Given the increase in qubit overhead, scaling the code in this way is only `worthwhile' if the resultant larger code has a lower logical error rate.

The \textit{threshold theorem} for stabilizer codes states that increasing the distance of a code will result in a corresponding reduction in the logical error rate $p_L$, provided the physical error rate $p$ of the individual code qubits is below a threshold $p<p_{th}$. The significance of this theorem is that it means that quantum error correction codes can in principle be used to arbitrarily suppress the logical error rate \cite{Preskill385,Kitaev1997,Aharonov:1997:FQC:258533.258579,Knill1998,Gottesman98}. Conversely, if the physical error rate is above the threshold, the process of quantum encoding becomes self defeating. The threshold of a code therefore provides a minimum experimental benchmark that quantum computing experiments must reach before quantum error correction becomes viable.

Upper bounds on the threshold $p_{th}$ for a code under a given noise model can be obtained using methods from statistical mechanics. Alternatively, more realistic thresholds can be numerically estimated by simulating code cycles and decoding using efficient inference algorithms as discussed in section \ref{sec:decode}. For the surface code, assuming $X$- and $Z$-errors are treated independently, the upper bound on the threshold is $\approx10.9\%$ \cite{Dennis02}. In practice, decoders based on the MWPM algorithm can achieve thresholds as high as $\approx 10.3\%$ \cite{Criger18}.

\subsection{Fault tolerance}

In the discussion of quantum error correction codes so far, we have assumed that errors only occur in certain locations in the circuit. For example, in the circuit diagram for the two-qubit code shown in figure \ref{fig:two_qubit}, errors are the restricted to the region labelled `$E$'. In doing this, we assume that all of the apparatus associated with encoding and syndrome extraction operates without error. However, in practice this is not the case. In fact, for many quantum computing technologies two-qubit gates, as well as measurement operations, can be dominant sources of error. As such, it is unrealistic to assume that any part of the circuit is error free.

A quantum error correction code is said to be \textit{fault tolerant} if it can account for errors (of size up to the code distance) that occur at any location in the circuit \cite{Gottesman98,Gottesman14}. Various techniques exist for modifying quantum circuits to make them fault tolerant \cite{Shor96,Steane97,DiVincenzo07}. In the simplest terms, these methods ensure that small sub-distance errors do not spread uncontrollably through the circuit.

Modifying a quantum error correction circuit for fault tolerance can add considerable overhead in terms of the total number of additional ancilla qubits required. For example, a fault tolerant syndrome extraction procedure proposed by Shor requires $\lambda$ ancilla qubits to measure each stabilizer, where $\lambda$ is the number of non-identity elements in the stabilizer \cite{Shor96}. Under this scheme, eight ancilla qubits would be required to measure the two stabilizers of the four-qubit code depicted in figure \ref{fig:four_qubit}. More efficient schemes exist \cite{Chao2018}, but a fault tolerant version of a code will always have increased overhead relative to the original circuit. 

For a quantum circuit with noisy ancilla measurements, it is not always possible to decode the error correction code in a single round of syndrome extraction. To illustrate this, consider  the case where the $S=10$ syndrome is measured in the three-qubit code outlined in section \ref{sec:three_qubit}. From table \ref{tab:three_qubit}, we see that this syndrome is triggered by an $X_1$ error. However, if the ancillas themselves are subject to error, then the same syndrome could equally have resulted from an error on ancilla qubit $A_1$. To differentiate between these two possibilities, it is necessary to perform two (or more) rounds of stabilizer measurements and compare the syndromes over time. It should be noted that decoding over time in this way, in addition to any other modifications required for fault tolerance, will reduce the threshold for the code. For example, the threshold for the surface code with noisy ancilla measurements is $\approx1\%$, compared to $\approx10\%$ in the ideal case \cite{Wang10,Fowler12}.

\subsection{Encoded computation}

A \textit{universal quantum computer} is a device that can perform any unitary operation $U$ that evolves a qubit register from one state to another $U\ket{\psi}=\ket{\psi'}$. It has been shown that any such operation $U$ can be efficiently compiled from a finite set of elementary gates \cite{Dawson05}. An example of a \textit{universal gate set} is $\langle \mathcal{U} \rangle=\langle  X, Z, Y, H, \cnoteq, T\rangle$, where $T={\rm diag} ( 1,e^{\rm i\pi/4} )$.

In this review, we have seen how Pauli logical $\bar{X}$ and $\bar{Z}$ operators can be defined for a variety of stabilizer codes. These gates allow computation to be performed directly on the encoded logical states, removing the need to decode then re-encode every time the state is to be evolved. However, the $\bar{X}$ and $\bar{Z}$ logical gates do not alone form a universal set, and additional logical operators need to be defined to achieve arbitrary computation on encoded states.

The major challenge in constructing a universal encoded gate set is to find ways in which the relevant gates can be performed fault tolerantly. For many codes, it is possible to fault tolerantly implement a subset of the gates in $\langle \mathcal{U} \rangle$ without having to introduce additional qubits. This is achieved by defining the logical operators with a property known as \textit{tranversality} that guarantees errors will not spread uncontrollably through the circuit. However, a no-go theorem exists that prohibits the implementation of a full universal gate set in this way on a quantum computer \cite{Eastin09}. As such, alternative techniques are required to perform universal encoded logic. Various methods have been proposed \cite{Bravyi05,Landahl14,Yoder17,Vasmer18}, but these typically impose a high cost in terms of the number of additional qubits required. To put this into context, it has proposed that the surface code could realise a universal gate set using a method called magic state injection \cite{Bravyi05}. However, estimates suggest that the fault tolerant implementation of this technique could result in an order-of-magnitude increase in the total number of qubits required in the quantum computer \cite{Campbell17}.

\subsection{Experimental implementations of quantum error correction}
The realisation of the first fault tolerant logical qubit will mark an important milestone in the journey to build a quantum computer. To this end, laboratories at places such as Google \cite{Kelly16}, IBM Research \cite{Gambetta2017,Takita2017} and TU Delft \cite{Rist2015} are currently building superconducting devices with the long-term goal of realising a surface code logical qubit. Other efforts are currently underway pursuing qubit architectures based on ion-trap trap technology \cite{Nickerson14} and quantum optics \cite{Qiang18}.

The threshold for the surface code under a realistic noise assumptions is approximately $1\%$ \cite{Wang10}. State-of-the-art qubit hardware has already been demonstrated with error rates below this level \cite{Chow12,Harty14}. However, suppressing the logical error rate to the point where the logical qubit outperforms an un-encoded qubit will require levels of scaleability that are not yet possible with current experiments. It is predicted that the first fault tolerant surface code logical qubits will require a lattice with over a thousand qubits \cite{Fowler12}. To put the scale of the challenge that remains into context, the largest quantum computers to date have less than one hundred qubits. Furthermore, achieving this goal will only be the first step: a quantum computer with only a single logical qubit will be no more powerful than an abacus with one bead. In fact, it is currently estimated that a fault tolerant surface code quantum computer with the ability to outperform a classical device for a \textit{useful} task will require over a million qubits in total \cite{OGorman17,Campbell17}.

The first quantum protocols to achieve fault tolerance will likely be quantum detection codes. As the smallest code capable of protecting against a quantum error model, the $[[4,2,2]]$ code is a promising candidate. Several proof-of-concept implementations of the $[[4,2,2]]$ code have already been demonstrated in \cite{Linke17,Vuillot18,Roffe18,Harper19}. Repetition codes (from the same family as the two- and three-qubit codes outlined in this review) have also been implemented on qubit hardware \cite{Wootton2018}. Over the past couple of years, several quantum computing hardware projects have developed cloud platforms to allow the public to program their devices. This makes it is possible for interested readers to test some of the early proof-of-concept quantum codes. For example, a tutorial showing how a repetition code can be implemented on the IBM Q device can be found in the supplementary material of \cite{Wootton2018}.    
  
\section{Outlook \& Summary}\label{sec:outlook}
A major hurdle in the realisation of a full-scale quantum computer stems from the challenge  of controlling qubits in an error free way. Quantum error correction protocols offer a solution to this problem, in principle allowing for arbitrary suppression of the logical error rate provided certain threshold conditions on the physical qubits are met. However, there is a trade-off: quantum error correction protocols require large number of qubits to operate effectively. This will significantly increase the overheads associated with quantum computing.

Developing quantum codes is not straightforward. Complications arise due the no-cloning theorem, the problem of wavefunction collapse and necessity to deal with multiple error types. Stabilizer codes provide a formalism that allow quantum error correction codes to be constructed within these constraints. For stabilizer codes, quantum redundancy is achieved by entangling the quantum information in the initial register across an expanded space of qubits. Errors can then be detected by performing a series of projective stabilizer measurements, and the results interpreted to determine the best recovery operation to restore the quantum information to its intended state.

The surface code is currently the most widely pursued quantum error correction scheme for experiment. This is due to its comparatively high threshold combined with the fact it requires only nearest-neighbour interactions. However, there are drawbacks to the surface code, most notably its poor encoding density. The distance of the surface code can be increased simply by scaling the size of the qubit lattice, but this results in a vanishing code rate, where the rate is defined as the ratio of encoded qubits to physical qubits $R=k/n$. Another disadvantage to the surface code is that resource intensive methods are required to obtain a universal encoded gate set. 

Alternatives to the surface code have been proposed based on different tilings of the qubit lattice \cite{Bombin2007}, as well as extensions to higher dimensions \cite{Breuckmann16,Vasmer18}. These constructions typically have lower thresholds, but offer other advantages such as (potentially) easier access to universal encoded gate sets \cite{Vasmer18}. Efforts are also in progress to develop code constructions with non-vanishing rates based on principles from high-performance classical codes \cite{Gottesman14,Tillich14}. However, for these codes, it is often necessary to perform arbitrary long range interactions between the code qubits.

In the quest to build a circuit model quantum computer there are many challenges to be overcome. At the hardware level, methods for the realisation and control of qubits need to be improved. In addition to this, a major theoretical challenge lies in finding better ways to achieve fault tolerant error correction. These two problems will have to be approached in parallel, with advances in either influencing the direction of the other.

\section*{Acknowledgement(s)}

JR acknowledges the support of the QCDA project which has received funding from the QuantERA ERA-NET Cofund in Quantum Technologies implemented within the European Union’s Horizon 2020 Programme. Many thanks to Viv Kendon for useful discussions and help preparing and checking the manuscript. Special thanks to Benjamin Jones and Armanda Ottaviano Quintavalle for reading through eary versions of the manuscript and providing valuable suggestions for improvement. Thanks also to Earl Campbell and Yingkai Ouyang for useful discussions, as well as to Jasminder Sidhu for helping with tikz drawings. The quantum circuit diagrams in this review were drawn using the QPIC package \cite{qpic}.

\section*{Notes on contributor(s)}

Joschka Roffe studied MPhys physics at The University Manchester, graduating in 2015. Following this, he studied a PhD in Quantum Computing at Durham University under the supervision of Viv Kendon. He now works as a research associate at the The University of Sheffield as part of the Quantum Codes Designs and Architectures (QCDA) project.

%


\bibliographystyle{tfnlm}
\bibliography{cp_review_arxiv}

\appendix

\section{Notation for quantum states}\label{sec:notation}

In this review quantum states are represented using Dirac bra-ket notation. Unless otherwise stated, we use the computational basis, given by $\{\ket{0},\ket{1} \}$ in the single-qubit case. For example, the general qubit state is written \begin{equation}
\ket{\psi}=\alpha\ket{0}+\beta\ket{1}\rm.
\end{equation}

For multi-qubit systems, we adopt a labelling convention whereby qubits are implicitly labelled $1,..,n$ from left-to-right, where $n$ is the total number of qubits. For example, the three-qubit basis element $\ket{010}$ is equivalent to $\ket{0}_1\otimes \ket{1}_2 \otimes \ket{0}_3$ in its full tensor product form.

\section{Pauli operator notation}\label{app:pauli}

The Pauli group on a single-qubit, $\mathcal{G}_1$, is defined as the set of Pauli operators 
\begin{equation}
\mathcal{G}_1=\{\pm \openone, \pm \text{i} \openone, \pm X, \pm \text{i} X, \pm Y, \pm \text{i} Y, \pm Z, \pm \text{i} Z \}\rm,
\end{equation}
where the $\pm 1$ and $\pm \text{i}$ terms are included to ensure $\mathcal{G}_1$ is closed under multiplication and thus forms a legitimate group. In matrix form, the four Pauli operators are given by
\begin{equation}
\openone = \left(\begin{matrix}
1 & 0\\0& 1
\end{matrix}\right), \quad
X=\left(\begin{matrix}
0 & 1\\1& 0
\end{matrix}\right), \quad
Y=\left(\begin{matrix}
0 & -\text{i}\\ \text{i}& 0
\end{matrix}\right), \quad
Z = \left(\begin{matrix}
1 & 0\\0& -1
\end{matrix}\right)\rm .
\end{equation}
The general Pauli group, $\mathcal{G}$, consists of the set of all operators that are formed from tensor products of the matrices in $\mathcal{G}_1$. For example, the operator
\begin{equation}\label{eq:pauli_op_example}
\openone \otimes X \otimes \openone \otimes Y \in \mathcal{G}
\end{equation}
is an element of the four-qubit Pauli group. The \textit{support} of a Pauli operator is given by the list of its non-identity elements. For example, the support of the Pauli operator in equation (\ref{eq:pauli_op_example}) is $X_2Y_4$ where the indices point to the qubit each element acts on. In this review, Pauli errors are always written in terms of their support. As an example, we would say that the bit-flip error $X_2$ acts on the two-qubit basis element $\ket{00}$ as follows $X_2\ket{00}=\ket{01}$.  

\section{Quantum circuit notation} \label{app:circuit_notation}

Quantum circuit notation provides a useful way of representing quantum algorithms. This appendix introduces the basic elements of quantum circuit notation necessary to understand quantum error correction circuits.

\subsection{Single qubit gates}

In quantum circuit representation of quantum algorithms each qubit in the quantum register is assigned a wire. These wires are labelled with quantum gates from left-to-right in the order in which they are applied during the quantum computation. As an example, consider the quantum computation described by the application of the unitary operation $U=X_1Z_1$ to a general qubit state,
\begin{equation}
\ket{\psi}=\alpha\ket{0}+\beta\ket{1} \xrightarrow{U=X_1Z_1}  X_1Z_1 \ket{\psi} = \alpha \ket{1}-\beta\ket{0}\rm.
\end{equation}
The quantum circuit for the above computation on a single-qubit is shown below
$$
\vcenter{\hbox{
		\resizebox{!}{!}{\input{figs/qec/xz.tikz}}}}\rm,
$$
where the input state is on the left and the output on the right. Note that the $Z$-gate is placed before the $X$-gate as it is applied first.

An important single-qubit gate for quantum error correction (and quantum algorithms in general) is the Hadamard gate which is defined in matrix-form as
\begin{equation}
H=\left(\begin{matrix}
1&1\\1&-1
\end{matrix}\right)\rm.
\end{equation}
The Hadamard gate has the following effect on the computational basis states
\begin{equation}
\begin{split}
H\ket{0}=\frac{1}{\sqrt{2}}(\ket{0}+\ket{1}),\\
H\ket{1}=\frac{1}{\sqrt{2}}(\ket{0}-\ket{1}).
\end{split}
\end{equation}

\subsection{Multi-qubit gates}

A gate spanning two wires in a quantum circuit represents a multi-qubit operation. As an example, the quantum circuit for the operation $U=X_1X_2$ applied the state $\ket{\psi}=\ket{00}$ is given by
$$
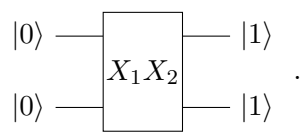
\vcenter{\hbox{
		\resizebox{!}{!}{\input{figs/qec/multi.tikz}}}}\rm.
$$

\subsection{Controlled-gates}

A controlled gate is a gate whose action is conditional on the value of a `control' qubit. In general, controlled gates are represented as follows in quantum circuit notation   
$$
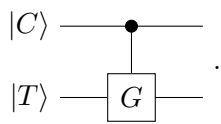
\vcenter{\hbox{
		\resizebox{!}{!}{\input{figs/qec/control.tikz}}}}\rm.
$$
In the above circuit, the top qubit $C$ is the control and the lower qubit $T$ is the target. The single-qubit $G$-gate is applied to the target if the control qubit is set to $\ket{C}=\ket{1}$. If the control qubit is set to $\ket{C}=0$, the $G$-gate is not applied to the target.

A commonly occurring gate in quantum error correction is the controlled-NOT (\cnot) gate. In this review, we use the two equivalent symbols for the \cnot gate
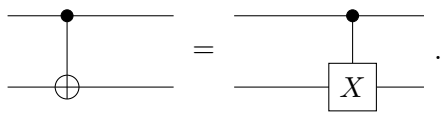
$$
\vcenter{\hbox{
		\resizebox{!}{!}{\input{figs/qec/cnot_equiv.tikz}}}}\rm.
$$

\subsection{Measurement in the computation basis}

In all the circuits in this review measurement is performed in the computation basis. As an example, consider the following quantum circuit for generating random numbers (the `Hello World' of quantum computing)
$$
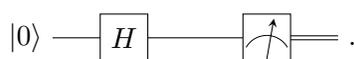
\vcenter{\hbox{
		\resizebox{!}{!}{\input{figs/qec/hello_world.tikz}}}}\rm.
$$
where the computational basis measurement is depicted by the gate to the right. The above circuit outputs `0' or `1' with equal probability. The double lines at the end of the circuit indicate that the output is classical information.

\section{Commutation properties for Pauli operators}\label{app:pauli_commute}

The elements of the Pauli group have eigenvalues $\{\pm 1, \pm \text{i}\}$. As a result, Pauli errors either commute or anti-commute with one another. In this appendix, we outline how to determine whether Pauli operators commute with one another.

First, recall that two operators, $F_i$ and $F_j$, commute if $F_iF_j=F_jF_i$, and anti-commute if $F_iF_j=(-1)F_jF_i$. For single-qubit Pauli-operators, we see that all pairs of distinct operators anti-commute
\begin{equation}
X_1Z_1=-Z_1X_1, \quad X_1Y_1=-Y_1X_1, \quad Z_1Y_1=-Y_1Z_1\rm. 	 
\end{equation}
Now consider the operators $Z_1Z_2$ and $X_1X_2$. These multi-qubit operators commute as
\begin{equation}
Z_1Z_2X_1X_2=Z_1X_1Z_2X_2=(-1)X_1Z_1(-1)X_2Z_2=X_1X_2Z_1Z_2\rm.
\end{equation}
In general, two Pauli operators will commute with one another if they intersect non-trivially on an even number of qubits as above. Conversely, if the number of non-trivial intersections is odd, then the two operators anti-commute. As an example, consider the two operators $X_1Z_2Z_3Z_5X_7$ and $X_1X_2X_5Z_7$. These operators intersect on the qubits $1$, $2$, $5$ and $7$. However, the intersection on qubit $1$ is trivial as both operators apply the same $X$-gate to that qubit. The number of non-trivial intersections that remain is therefore three. As the two operators intersect non-trivially an odd number of times, the two operators anti-commute.

\end{document}

%% file: bloch_sphere/bloch.tikz

\begin{tikzpicture}[line cap=round, line join=round, >=Triangle]
  \clip(-2.1,-2.1) rectangle (2.38,2.58);

  \draw[ball color=gray!20!white, fill opacity=0.6] (0,0) circle (2cm);
  \draw [rotate around={0.:(0.,0.)},dash pattern=on 3pt off 3pt] (0,0) ellipse (2cm and 0.9cm);
  \draw (0,0)-- (0.70,1.07);

  \draw[-to,line width=0.7pt]   (0,0) -- +(0,2);
  \draw[-to,line width=0.7pt]   (0,0) -- +(-0.83,-0.81);
  \draw[-to,line width=0.7pt]   (0,0) -- +(2,0);    

  \draw [-Latex, <-, >=stealth', shift={(0,0)}, black, fill opacity=1] (56.7:0.4) arc (56.7:90.:0.4);
  
  \begin{scope}[rotate around x=10, y=10, xshift=1, yshift=-4.6]
  \draw [-Latex, ->, >=stealth', shift={(0,0)}, black, fill opacity=1] (-135.7:0.4) arc (-135.7:-33.2:0.4);
  \end{scope}
  
  \draw [dotted] (0.7,1)-- (0.7,-0.46);
  \draw [dotted] (0,0)-- (0.7,-0.46);
  
  \draw (-0.2,-0.29) node[anchor=north west] {$\phi$};
  \draw (-0.08,0.9) node[anchor=north west] {$\theta$};
  \draw (-1.15,-0.75) node[anchor=north west] {\small{$\hat{\bm{x}}$}};
  \draw (1.95,0.3) node[anchor=north west] {\small{$\hat{\bm{y}}$}};
  \draw (-0.5,2.6) node[anchor=north west] {\small{$\hat{\bm{z}}=\ket{0}$}};
  \draw (0.4,1.65) node[anchor=north west] {$\ket{\psi}$};

  \draw [fill] (0,0) circle (1.5pt);
  \draw [fill] (0.7,1.07) circle (0.5pt);
\end{tikzpicture}

%% file: figs/qec/two_qubit_v2.tikz
\usetikzlibrary{decorations.pathreplacing,decorations.pathmorphing}
\providecommand{\ket}[1]{\left|#1\right\rangle}
\begin{tikzpicture}[scale=1.500000,x=1pt,y=1pt]
\filldraw[color=white] (0.000000, -12.000000) rectangle (185.000000, 60.000000);
\draw[color=black] (0.000000,48.000000) -- (185.000000,48.000000);
\draw[color=black] (0.000000,48.000000) node[left] {$\ket{\psi}_1$};
\draw[color=black] (0.000000,24.000000) -- (185.000000,24.000000);
\draw[color=black] (0.000000,24.000000) node[left] {$\ket{0}_2$};
\draw[color=black] (88.500000,0.000000) -- (174.500000,0.000000);
\draw[color=black] (174.500000,-0.500000) -- (185.000000,-0.500000);
\draw[color=black] (174.500000,0.500000) -- (185.000000,0.500000);
\draw (12.000000,48.000000) -- (12.000000,24.000000);
\filldraw (12.000000, 48.000000) circle(1.500000pt);
\begin{scope}
\draw[fill=white] (12.000000, 24.000000) circle(3.000000pt);
\clip (12.000000, 24.000000) circle(3.000000pt);
\draw (9.000000, 24.000000) -- (15.000000, 24.000000);
\draw (12.000000, 21.000000) -- (12.000000, 27.000000);
\end{scope}
\draw[fill=white,color=white] (21.000000, 18.000000) rectangle (36.000000, 54.000000);
\draw (28.500000, 36.000000) node {$\ket{\psi}_L$};
\draw (48.000000,48.000000) -- (48.000000,24.000000);
\begin{scope}
\draw[fill=white] (48.000000, 36.000000) +(-45.000000:8.485281pt and 25.455844pt) -- +(45.000000:8.485281pt and 25.455844pt) -- +(135.000000:8.485281pt and 25.455844pt) -- +(225.000000:8.485281pt and 25.455844pt) -- cycle;
\clip (48.000000, 36.000000) +(-45.000000:8.485281pt and 25.455844pt) -- +(45.000000:8.485281pt and 25.455844pt) -- +(135.000000:8.485281pt and 25.455844pt) -- +(225.000000:8.485281pt and 25.455844pt) -- cycle;
\draw (48.000000, 36.000000) node {$E$};
\end{scope}
\draw[fill=white,color=white] (60.000000, 18.000000) rectangle (75.000000, 54.000000);
\draw (67.500000, 36.000000) node {$E\ket{\psi}_L$};
\draw[color=black] (96.000000,0.000000) node[fill=white,left,minimum height=24.000000pt,minimum width=15.000000pt,inner sep=0pt] {\phantom{$\ket{0}_A$}};
\draw[color=black] (96.000000,0.000000) node[left] {$\ket{0}_A$};
\begin{scope}
\draw[fill=white] (108.000000, -0.000000) +(-45.000000:8.485281pt and 8.485281pt) -- +(45.000000:8.485281pt and 8.485281pt) -- +(135.000000:8.485281pt and 8.485281pt) -- +(225.000000:8.485281pt and 8.485281pt) -- cycle;
\clip (108.000000, -0.000000) +(-45.000000:8.485281pt and 8.485281pt) -- +(45.000000:8.485281pt and 8.485281pt) -- +(135.000000:8.485281pt and 8.485281pt) -- +(225.000000:8.485281pt and 8.485281pt) -- cycle;
\draw (108.000000, -0.000000) node {$H$};
\end{scope}
\draw (131.500000,48.000000) -- (131.500000,0.000000);
\begin{scope}
\draw[fill=white] (131.500000, 36.000000) +(-45.000000:16.263456pt and 25.455844pt) -- +(45.000000:16.263456pt and 25.455844pt) -- +(135.000000:16.263456pt and 25.455844pt) -- +(225.000000:16.263456pt and 25.455844pt) -- cycle;
\clip (131.500000, 36.000000) +(-45.000000:16.263456pt and 25.455844pt) -- +(45.000000:16.263456pt and 25.455844pt) -- +(135.000000:16.263456pt and 25.455844pt) -- +(225.000000:16.263456pt and 25.455844pt) -- cycle;
\draw (131.500000, 36.000000) node {$Z_1 Z_2$};
\end{scope}
\filldraw (131.500000, 0.000000) circle(1.500000pt);
\begin{scope}
\draw[fill=white] (155.000000, -0.000000) +(-45.000000:8.485281pt and 8.485281pt) -- +(45.000000:8.485281pt and 8.485281pt) -- +(135.000000:8.485281pt and 8.485281pt) -- +(225.000000:8.485281pt and 8.485281pt) -- cycle;
\clip (155.000000, -0.000000) +(-45.000000:8.485281pt and 8.485281pt) -- +(45.000000:8.485281pt and 8.485281pt) -- +(135.000000:8.485281pt and 8.485281pt) -- +(225.000000:8.485281pt and 8.485281pt) -- cycle;
\draw (155.000000, -0.000000) node {$H$};
\end{scope}
\draw[fill=white,color=white] (167.000000, 18.000000) rectangle (182.000000, 54.000000);
\draw (174.500000, 36.000000) node {$E\ket{\psi}_L$};
\draw[fill=white] (168.500000, -6.000000) rectangle (180.500000, 6.000000);
\draw[very thin] (174.500000, 0.600000) arc (90:150:6.000000pt);
\draw[very thin] (174.500000, 0.600000) arc (90:30:6.000000pt);
\draw[->,>=stealth] (174.500000, -5.400000) -- +(80:10.392305pt);
\draw[decorate,decoration={brace,amplitude = 2.250000pt},very thick] (7.500000,60.000000) -- (16.500000,60.000000);
\draw (12.000000, 64.000000) node[text width=144pt,above,text centered] {Encoder};
\draw[decorate,decoration={brace,amplitude = 4.000000pt},very thick] (100.500000,60.000000) -- (162.500000,60.000000);
\draw (131.500000, 64.000000) node[text width=144pt,above,text centered] {{Syndrome extraction}};
\end{tikzpicture}

%% file: figs/qec/three_qubit.tikz
\providecommand{\ket}[1]{\left|#1\right\rangle}
\begin{tikzpicture}[scale=1.500000,x=1pt,y=1pt]
\filldraw[color=white] (0.000000, -9.000000) rectangle (194.000000, 81.000000);
\draw[color=black] (0.000000,72.000000) -- (194.000000,72.000000);
\draw[color=black] (0.000000,72.000000) node[left] {$\ket{\psi}_1$};
\draw[color=black] (0.000000,54.000000) -- (194.000000,54.000000);
\draw[color=black] (0.000000,54.000000) node[left] {$\ket{0}_2$};
\draw[color=black] (0.000000,36.000000) -- (194.000000,36.000000);
\draw[color=black] (0.000000,36.000000) node[left] {$\ket{0}_3$};
\draw[color=black] (71.500000,18.000000) -- (185.000000,18.000000);
\draw[color=black] (185.000000,17.500000) -- (194.000000,17.500000);
\draw[color=black] (185.000000,18.500000) -- (194.000000,18.500000);
\draw[color=black] (71.500000,0.000000) -- (185.000000,0.000000);
\draw[color=black] (185.000000,-0.500000) -- (194.000000,-0.500000);
\draw[color=black] (185.000000,0.500000) -- (194.000000,0.500000);
\draw (12.000000,72.000000) -- (12.000000,54.000000);
\filldraw (12.000000, 72.000000) circle(1.500000pt);
\begin{scope}
\draw[fill=white] (12.000000, 54.000000) circle(3.000000pt);
\clip (12.000000, 54.000000) circle(3.000000pt);
\draw (9.000000, 54.000000) -- (15.000000, 54.000000);
\draw (12.000000, 51.000000) -- (12.000000, 57.000000);
\end{scope}
\draw (24.000000,54.000000) -- (24.000000,36.000000);
\filldraw (24.000000, 54.000000) circle(1.500000pt);
\begin{scope}
\draw[fill=white] (24.000000, 36.000000) circle(3.000000pt);
\clip (24.000000, 36.000000) circle(3.000000pt);
\draw (21.000000, 36.000000) -- (27.000000, 36.000000);
\draw (24.000000, 33.000000) -- (24.000000, 39.000000);
\end{scope}
\draw[fill=white,color=white] (33.000000, 30.000000) rectangle (58.000000, 78.000000);
\draw (45.500000, 54.000000) node {$\ket{\psi}_L$};
\draw[color=black] (79.000000,18.000000) node[fill=white,left,minimum height=18.000000pt,minimum width=15.000000pt,inner sep=0pt] {\phantom{$\ket{0}_{A_1}$}};
\draw[color=black] (79.000000,18.000000) node[left] {$\ket{0}_{A_1}$};
\draw[color=black] (79.000000,0.000000) node[fill=white,left,minimum height=18.000000pt,minimum width=15.000000pt,inner sep=0pt] {\phantom{$\ket{0}_{A_2}$}};
\draw[color=black] (79.000000,0.000000) node[left] {$\ket{0}_{A_2}$};
\draw (71.500000,72.000000) -- (71.500000,36.000000);
\begin{scope}
\draw[fill=white] (71.500000, 54.000000) +(-45.000000:8.485281pt and 33.941125pt) -- +(45.000000:8.485281pt and 33.941125pt) -- +(135.000000:8.485281pt and 33.941125pt) -- +(225.000000:8.485281pt and 33.941125pt) -- cycle;
\clip (71.500000, 54.000000) +(-45.000000:8.485281pt and 33.941125pt) -- +(45.000000:8.485281pt and 33.941125pt) -- +(135.000000:8.485281pt and 33.941125pt) -- +(225.000000:8.485281pt and 33.941125pt) -- cycle;
\draw (71.500000, 54.000000) node {$E$};
\end{scope}
\begin{scope}
\draw[fill=white] (91.000000, 18.000000) +(-45.000000:8.485281pt and 8.485281pt) -- +(45.000000:8.485281pt and 8.485281pt) -- +(135.000000:8.485281pt and 8.485281pt) -- +(225.000000:8.485281pt and 8.485281pt) -- cycle;
\clip (91.000000, 18.000000) +(-45.000000:8.485281pt and 8.485281pt) -- +(45.000000:8.485281pt and 8.485281pt) -- +(135.000000:8.485281pt and 8.485281pt) -- +(225.000000:8.485281pt and 8.485281pt) -- cycle;
\draw (91.000000, 18.000000) node {$H$};
\end{scope}
\begin{scope}
\draw[fill=white] (91.000000, -0.000000) +(-45.000000:8.485281pt and 8.485281pt) -- +(45.000000:8.485281pt and 8.485281pt) -- +(135.000000:8.485281pt and 8.485281pt) -- +(225.000000:8.485281pt and 8.485281pt) -- cycle;
\clip (91.000000, -0.000000) +(-45.000000:8.485281pt and 8.485281pt) -- +(45.000000:8.485281pt and 8.485281pt) -- +(135.000000:8.485281pt and 8.485281pt) -- +(225.000000:8.485281pt and 8.485281pt) -- cycle;
\draw (91.000000, -0.000000) node {$H$};
\end{scope}
\draw (114.500000,72.000000) -- (114.500000,18.000000);
\begin{scope}
\draw[fill=white] (114.500000, 63.000000) +(-45.000000:16.263456pt and 21.213203pt) -- +(45.000000:16.263456pt and 21.213203pt) -- +(135.000000:16.263456pt and 21.213203pt) -- +(225.000000:16.263456pt and 21.213203pt) -- cycle;
\clip (114.500000, 63.000000) +(-45.000000:16.263456pt and 21.213203pt) -- +(45.000000:16.263456pt and 21.213203pt) -- +(135.000000:16.263456pt and 21.213203pt) -- +(225.000000:16.263456pt and 21.213203pt) -- cycle;
\draw (114.500000, 63.000000) node {$Z_1 Z_2$};
\end{scope}
\filldraw (114.500000, 18.000000) circle(1.500000pt);
\draw (143.500000,54.000000) -- (143.500000,0.000000);
\begin{scope}
\draw[fill=white] (143.500000, 45.000000) +(-45.000000:16.263456pt and 21.213203pt) -- +(45.000000:16.263456pt and 21.213203pt) -- +(135.000000:16.263456pt and 21.213203pt) -- +(225.000000:16.263456pt and 21.213203pt) -- cycle;
\clip (143.500000, 45.000000) +(-45.000000:16.263456pt and 21.213203pt) -- +(45.000000:16.263456pt and 21.213203pt) -- +(135.000000:16.263456pt and 21.213203pt) -- +(225.000000:16.263456pt and 21.213203pt) -- cycle;
\draw (143.500000, 45.000000) node {$Z_2Z_3$};
\end{scope}
\filldraw (143.500000, 0.000000) circle(1.500000pt);
\begin{scope}
\draw[fill=white] (167.000000, -0.000000) +(-45.000000:8.485281pt and 8.485281pt) -- +(45.000000:8.485281pt and 8.485281pt) -- +(135.000000:8.485281pt and 8.485281pt) -- +(225.000000:8.485281pt and 8.485281pt) -- cycle;
\clip (167.000000, -0.000000) +(-45.000000:8.485281pt and 8.485281pt) -- +(45.000000:8.485281pt and 8.485281pt) -- +(135.000000:8.485281pt and 8.485281pt) -- +(225.000000:8.485281pt and 8.485281pt) -- cycle;
\draw (167.000000, -0.000000) node {$H$};
\end{scope}
\begin{scope}
\draw[fill=white] (167.000000, 18.000000) +(-45.000000:8.485281pt and 8.485281pt) -- +(45.000000:8.485281pt and 8.485281pt) -- +(135.000000:8.485281pt and 8.485281pt) -- +(225.000000:8.485281pt and 8.485281pt) -- cycle;
\clip (167.000000, 18.000000) +(-45.000000:8.485281pt and 8.485281pt) -- +(45.000000:8.485281pt and 8.485281pt) -- +(135.000000:8.485281pt and 8.485281pt) -- +(225.000000:8.485281pt and 8.485281pt) -- cycle;
\draw (167.000000, 18.000000) node {$H$};
\end{scope}
\draw[fill=white] (179.000000, 12.000000) rectangle (191.000000, 24.000000);
\draw[very thin] (185.000000, 18.600000) arc (90:150:6.000000pt);
\draw[very thin] (185.000000, 18.600000) arc (90:30:6.000000pt);
\draw[->,>=stealth] (185.000000, 12.600000) -- +(80:10.392305pt);
\draw[fill=white] (179.000000, -6.000000) rectangle (191.000000, 6.000000);
\draw[very thin] (185.000000, 0.600000) arc (90:150:6.000000pt);
\draw[very thin] (185.000000, 0.600000) arc (90:30:6.000000pt);
\draw[->,>=stealth] (185.000000, -5.400000) -- +(80:10.392305pt);
\draw[decorate,decoration={brace,amplitude = 4.000000pt},very thick] (7.500000,81.000000) -- (28.500000,81.000000);
\draw (18.000000, 85.000000) node[text width=144pt,above,text centered] {Encoder};
\draw[decorate,decoration={brace,amplitude = 4.000000pt},very thick] (83.500000,81.000000) -- (174.500000,81.000000);
\draw (129.000000, 85.000000) node[text width=144pt,above,text centered] {{Syndrome extraction}};
\end{tikzpicture}

%% file: figs/qec/stab_codes.tikz
\usetikzlibrary{decorations.pathreplacing,decorations.pathmorphing}
\providecommand{\ket}[1]{\left|#1\right\rangle}
\begin{tikzpicture}[scale=1.500000,x=1pt,y=1pt]
\filldraw[color=white] (0.000000, -9.000000) rectangle (260.000000, 99.000000);
\draw[color=black] (0.000000,90.000000) -- (260.000000,90.000000);
\draw[color=black] (0.000000,90.000000) node[left] {$\ket{\psi}_D$};
\draw[color=black] (0.000000,72.000000) -- (260.000000,72.000000);
\draw[color=black] (0.000000,72.000000) node[left] {$\ket{0}_R$};
\draw[color=black] (96.500000,54.000000) -- (251.000000,54.000000);
\draw[color=black] (251.000000,53.500000) -- (260.000000,53.500000);
\draw[color=black] (251.000000,54.500000) -- (260.000000,54.500000);
\draw[color=black] (96.500000,36.000000) -- (251.000000,36.000000);
\draw[color=black] (251.000000,35.500000) -- (260.000000,35.500000);
\draw[color=black] (251.000000,36.500000) -- (260.000000,36.500000);
\draw[color=black] (96.500000,0.000000) -- (251.000000,0.000000);
\draw[color=black] (251.000000,-0.500000) -- (260.000000,-0.500000);
\draw[color=black] (251.000000,0.500000) -- (260.000000,0.500000);
\draw (3.000000, 84.000000) -- (11.000000, 96.000000);
\draw (3.000000, 66.000000) -- (11.000000, 78.000000);
\draw (34.500000,90.000000) -- (34.500000,72.000000);
\begin{scope}
\draw[fill=white] (34.500000, 81.000000) +(-45.000000:24.748737pt and 21.213203pt) -- +(45.000000:24.748737pt and 21.213203pt) -- +(135.000000:24.748737pt and 21.213203pt) -- +(225.000000:24.748737pt and 21.213203pt) -- cycle;
\clip (34.500000, 81.000000) +(-45.000000:24.748737pt and 21.213203pt) -- +(45.000000:24.748737pt and 21.213203pt) -- +(135.000000:24.748737pt and 21.213203pt) -- +(225.000000:24.748737pt and 21.213203pt) -- cycle;
\draw (34.500000, 81.000000) node {{Encoder}};
\end{scope}
\draw[fill=white,color=white] (58.000000, 66.000000) rectangle (83.000000, 96.000000);
\draw (70.500000, 81.000000) node {${\ket{\psi}_L}$};
\draw (96.500000,90.000000) -- (96.500000,72.000000);
\begin{scope}
\draw[fill=white] (96.500000, 81.000000) +(-45.000000:8.485281pt and 21.213203pt) -- +(45.000000:8.485281pt and 21.213203pt) -- +(135.000000:8.485281pt and 21.213203pt) -- +(225.000000:8.485281pt and 21.213203pt) -- cycle;
\clip (96.500000, 81.000000) +(-45.000000:8.485281pt and 21.213203pt) -- +(45.000000:8.485281pt and 21.213203pt) -- +(135.000000:8.485281pt and 21.213203pt) -- +(225.000000:8.485281pt and 21.213203pt) -- cycle;
\draw (96.500000, 81.000000) node {{$E$}};
\end{scope}
\draw[color=black] (104.000000,54.000000) node[fill=white,left,minimum height=18.000000pt,minimum width=15.000000pt,inner sep=0pt] {\phantom{$\ket{0}_{A_1}$}};
\draw[color=black] (104.000000,54.000000) node[left] {$\ket{0}_{A_1}$};
\draw[color=black] (104.000000,18.000000) node[fill=white,left,minimum height=18.000000pt,minimum width=15.000000pt,anchor = base,inner sep=0pt] {\phantom{$\;\vdots\;$}};
\draw[color=black] (104.000000,18.000000) node[anchor=mid east] {$\vdots$};
\draw[color=black] (104.000000,36.000000) node[fill=white,left,minimum height=18.000000pt,minimum width=15.000000pt,inner sep=0pt] {\phantom{$\ket{0}_{A_2}$}};
\draw[color=black] (104.000000,36.000000) node[left] {$\ket{0}_{A_2}$};
\draw[color=black] (104.000000,0.000000) node[fill=white,left,minimum height=18.000000pt,minimum width=15.000000pt,inner sep=0pt] {\phantom{$\ket{0}_{A_{n-k}}$}};
\draw[color=black] (104.000000,0.000000) node[left] {$\ket{0}_{A_{n-k}}$};
\begin{scope}
\draw[fill=white] (116.000000, 54.000000) +(-45.000000:8.485281pt and 8.485281pt) -- +(45.000000:8.485281pt and 8.485281pt) -- +(135.000000:8.485281pt and 8.485281pt) -- +(225.000000:8.485281pt and 8.485281pt) -- cycle;
\clip (116.000000, 54.000000) +(-45.000000:8.485281pt and 8.485281pt) -- +(45.000000:8.485281pt and 8.485281pt) -- +(135.000000:8.485281pt and 8.485281pt) -- +(225.000000:8.485281pt and 8.485281pt) -- cycle;
\draw (116.000000, 54.000000) node {$H$};
\end{scope}
\begin{scope}
\draw[fill=white] (116.000000, 36.000000) +(-45.000000:8.485281pt and 8.485281pt) -- +(45.000000:8.485281pt and 8.485281pt) -- +(135.000000:8.485281pt and 8.485281pt) -- +(225.000000:8.485281pt and 8.485281pt) -- cycle;
\clip (116.000000, 36.000000) +(-45.000000:8.485281pt and 8.485281pt) -- +(45.000000:8.485281pt and 8.485281pt) -- +(135.000000:8.485281pt and 8.485281pt) -- +(225.000000:8.485281pt and 8.485281pt) -- cycle;
\draw (116.000000, 36.000000) node {$H$};
\end{scope}
\begin{scope}
\draw[fill=white] (116.000000, -0.000000) +(-45.000000:8.485281pt and 8.485281pt) -- +(45.000000:8.485281pt and 8.485281pt) -- +(135.000000:8.485281pt and 8.485281pt) -- +(225.000000:8.485281pt and 8.485281pt) -- cycle;
\clip (116.000000, -0.000000) +(-45.000000:8.485281pt and 8.485281pt) -- +(45.000000:8.485281pt and 8.485281pt) -- +(135.000000:8.485281pt and 8.485281pt) -- +(225.000000:8.485281pt and 8.485281pt) -- cycle;
\draw (116.000000, -0.000000) node {$H$};
\end{scope}
\draw (138.000000,90.000000) -- (138.000000,54.000000);
\begin{scope}
\draw[fill=white] (138.000000, 81.000000) +(-45.000000:14.142136pt and 21.213203pt) -- +(45.000000:14.142136pt and 21.213203pt) -- +(135.000000:14.142136pt and 21.213203pt) -- +(225.000000:14.142136pt and 21.213203pt) -- cycle;
\clip (138.000000, 81.000000) +(-45.000000:14.142136pt and 21.213203pt) -- +(45.000000:14.142136pt and 21.213203pt) -- +(135.000000:14.142136pt and 21.213203pt) -- +(225.000000:14.142136pt and 21.213203pt) -- cycle;
\draw (138.000000, 81.000000) node {{$P_1$}};
\end{scope}
\filldraw (138.000000, 54.000000) circle(1.500000pt);
\draw (164.000000,90.000000) -- (164.000000,36.000000);
\begin{scope}
\draw[fill=white] (164.000000, 81.000000) +(-45.000000:14.142136pt and 21.213203pt) -- +(45.000000:14.142136pt and 21.213203pt) -- +(135.000000:14.142136pt and 21.213203pt) -- +(225.000000:14.142136pt and 21.213203pt) -- cycle;
\clip (164.000000, 81.000000) +(-45.000000:14.142136pt and 21.213203pt) -- +(45.000000:14.142136pt and 21.213203pt) -- +(135.000000:14.142136pt and 21.213203pt) -- +(225.000000:14.142136pt and 21.213203pt) -- cycle;
\draw (164.000000, 81.000000) node {{$P_2$}};
\end{scope}
\filldraw (164.000000, 36.000000) circle(1.500000pt);
\draw[fill=white,color=white] (180.000000, 66.000000) rectangle (195.000000, 96.000000);
\draw (187.500000, 81.000000) node {$...$};
\draw (211.000000,90.000000) -- (211.000000,0.000000);
\begin{scope}
\draw[fill=white] (211.000000, 81.000000) +(-45.000000:14.142136pt and 21.213203pt) -- +(45.000000:14.142136pt and 21.213203pt) -- +(135.000000:14.142136pt and 21.213203pt) -- +(225.000000:14.142136pt and 21.213203pt) -- cycle;
\clip (211.000000, 81.000000) +(-45.000000:14.142136pt and 21.213203pt) -- +(45.000000:14.142136pt and 21.213203pt) -- +(135.000000:14.142136pt and 21.213203pt) -- +(225.000000:14.142136pt and 21.213203pt) -- cycle;
\draw (211.000000, 81.000000) node {{$P_{n-k}$}};
\end{scope}
\filldraw (211.000000, 0.000000) circle(1.500000pt);
\begin{scope}
\draw[fill=white] (233.000000, 54.000000) +(-45.000000:8.485281pt and 8.485281pt) -- +(45.000000:8.485281pt and 8.485281pt) -- +(135.000000:8.485281pt and 8.485281pt) -- +(225.000000:8.485281pt and 8.485281pt) -- cycle;
\clip (233.000000, 54.000000) +(-45.000000:8.485281pt and 8.485281pt) -- +(45.000000:8.485281pt and 8.485281pt) -- +(135.000000:8.485281pt and 8.485281pt) -- +(225.000000:8.485281pt and 8.485281pt) -- cycle;
\draw (233.000000, 54.000000) node {$H$};
\end{scope}
\begin{scope}
\draw[fill=white] (233.000000, 36.000000) +(-45.000000:8.485281pt and 8.485281pt) -- +(45.000000:8.485281pt and 8.485281pt) -- +(135.000000:8.485281pt and 8.485281pt) -- +(225.000000:8.485281pt and 8.485281pt) -- cycle;
\clip (233.000000, 36.000000) +(-45.000000:8.485281pt and 8.485281pt) -- +(45.000000:8.485281pt and 8.485281pt) -- +(135.000000:8.485281pt and 8.485281pt) -- +(225.000000:8.485281pt and 8.485281pt) -- cycle;
\draw (233.000000, 36.000000) node {$H$};
\end{scope}
\begin{scope}
\draw[fill=white] (233.000000, -0.000000) +(-45.000000:8.485281pt and 8.485281pt) -- +(45.000000:8.485281pt and 8.485281pt) -- +(135.000000:8.485281pt and 8.485281pt) -- +(225.000000:8.485281pt and 8.485281pt) -- cycle;
\clip (233.000000, -0.000000) +(-45.000000:8.485281pt and 8.485281pt) -- +(45.000000:8.485281pt and 8.485281pt) -- +(135.000000:8.485281pt and 8.485281pt) -- +(225.000000:8.485281pt and 8.485281pt) -- cycle;
\draw (233.000000, -0.000000) node {$H$};
\end{scope}
\draw[fill=white] (245.000000, 48.000000) rectangle (257.000000, 60.000000);
\draw[very thin] (251.000000, 54.600000) arc (90:150:6.000000pt);
\draw[very thin] (251.000000, 54.600000) arc (90:30:6.000000pt);
\draw[->,>=stealth] (251.000000, 48.600000) -- +(80:10.392305pt);
\draw[fill=white] (245.000000, 30.000000) rectangle (257.000000, 42.000000);
\draw[very thin] (251.000000, 36.600000) arc (90:150:6.000000pt);
\draw[very thin] (251.000000, 36.600000) arc (90:30:6.000000pt);
\draw[->,>=stealth] (251.000000, 30.600000) -- +(80:10.392305pt);
\draw[fill=white] (245.000000, -6.000000) rectangle (257.000000, 6.000000);
\draw[very thin] (251.000000, 0.600000) arc (90:150:6.000000pt);
\draw[very thin] (251.000000, 0.600000) arc (90:30:6.000000pt);
\draw[->,>=stealth] (251.000000, -5.400000) -- +(80:10.392305pt);
\draw[color=black] (260.000000,18.000000) node[anchor=mid west] {$\vdots$};
\draw[decorate,decoration={brace,amplitude = 4.000000pt},very thick] (108.500000,99.000000) -- (240.500000,99.000000);
\draw (174.500000, 103.000000) node[text width=144pt,above,text centered] {{Syndrome extraction}};
\end{tikzpicture}

%% file: figs/qec/422_code.tikz
\usetikzlibrary{decorations.pathreplacing,decorations.pathmorphing}
\providecommand{\ket}[1]{\left|#1\right\rangle}
\begin{tikzpicture}[scale=1.500000,x=1pt,y=1pt]
\filldraw[color=white] (0.000000, -9.000000) rectangle (241.000000, 99.000000);
\draw[color=black] (0.000000,90.000000) -- (241.000000,90.000000);
\draw[color=black] (0.000000,90.000000) node[left] {$\ket{\psi}_1$};
\draw[color=black] (0.000000,72.000000) -- (241.000000,72.000000);
\draw[color=black] (0.000000,72.000000) node[left] {$\ket{\psi}_2$};
\draw[color=black] (0.000000,54.000000) -- (241.000000,54.000000);
\draw[color=black] (0.000000,54.000000) node[left] {$\ket{0}_3$};
\draw[color=black] (0.000000,36.000000) -- (241.000000,36.000000);
\draw[color=black] (0.000000,36.000000) node[left] {$\ket{0}_4$};
\draw[color=black] (118.500000,18.000000) -- (232.000000,18.000000);
\draw[color=black] (232.000000,17.500000) -- (241.000000,17.500000);
\draw[color=black] (232.000000,18.500000) -- (241.000000,18.500000);
\draw[color=black] (118.500000,0.000000) -- (232.000000,0.000000);
\draw[color=black] (232.000000,-0.500000) -- (241.000000,-0.500000);
\draw[color=black] (232.000000,0.500000) -- (241.000000,0.500000);
\begin{scope}
\draw[fill=white] (15.000000, 36.000000) +(-45.000000:8.485281pt and 8.485281pt) -- +(45.000000:8.485281pt and 8.485281pt) -- +(135.000000:8.485281pt and 8.485281pt) -- +(225.000000:8.485281pt and 8.485281pt) -- cycle;
\clip (15.000000, 36.000000) +(-45.000000:8.485281pt and 8.485281pt) -- +(45.000000:8.485281pt and 8.485281pt) -- +(135.000000:8.485281pt and 8.485281pt) -- +(225.000000:8.485281pt and 8.485281pt) -- cycle;
\draw (15.000000, 36.000000) node {$H$};
\end{scope}
\draw (15.000000,90.000000) -- (15.000000,54.000000);
\filldraw (15.000000, 90.000000) circle(1.500000pt);
\begin{scope}
\draw[fill=white] (15.000000, 54.000000) circle(3.000000pt);
\clip (15.000000, 54.000000) circle(3.000000pt);
\draw (12.000000, 54.000000) -- (18.000000, 54.000000);
\draw (15.000000, 51.000000) -- (15.000000, 57.000000);
\end{scope}
\draw (30.000000,72.000000) -- (30.000000,54.000000);
\filldraw (30.000000, 72.000000) circle(1.500000pt);
\begin{scope}
\draw[fill=white] (30.000000, 54.000000) circle(3.000000pt);
\clip (30.000000, 54.000000) circle(3.000000pt);
\draw (27.000000, 54.000000) -- (33.000000, 54.000000);
\draw (30.000000, 51.000000) -- (30.000000, 57.000000);
\end{scope}
\draw (42.000000,54.000000) -- (42.000000,36.000000);
\filldraw (42.000000, 36.000000) circle(1.500000pt);
\begin{scope}
\draw[fill=white] (42.000000, 54.000000) circle(3.000000pt);
\clip (42.000000, 54.000000) circle(3.000000pt);
\draw (39.000000, 54.000000) -- (45.000000, 54.000000);
\draw (42.000000, 51.000000) -- (42.000000, 57.000000);
\end{scope}
\draw (54.000000,72.000000) -- (54.000000,36.000000);
\filldraw (54.000000, 36.000000) circle(1.500000pt);
\begin{scope}
\draw[fill=white] (54.000000, 72.000000) circle(3.000000pt);
\clip (54.000000, 72.000000) circle(3.000000pt);
\draw (51.000000, 72.000000) -- (57.000000, 72.000000);
\draw (54.000000, 69.000000) -- (54.000000, 75.000000);
\end{scope}
\draw (66.000000,90.000000) -- (66.000000,36.000000);
\filldraw (66.000000, 36.000000) circle(1.500000pt);
\begin{scope}
\draw[fill=white] (66.000000, 90.000000) circle(3.000000pt);
\clip (66.000000, 90.000000) circle(3.000000pt);
\draw (63.000000, 90.000000) -- (69.000000, 90.000000);
\draw (66.000000, 87.000000) -- (66.000000, 93.000000);
\end{scope}
\draw[fill=white,color=white] (75.000000, 30.000000) rectangle (105.000000, 96.000000);
\draw (90.000000, 63.000000) node {$\ket{\psi_1\psi_2}_L$};
\draw[color=black] (126.000000,18.000000) node[fill=white,left,minimum height=18.000000pt,minimum width=15.000000pt,inner sep=0pt] {\phantom{$\ket{0}_{A_1}$}};
\draw[color=black] (126.000000,18.000000) node[left] {$\ket{0}_{A_1}$};
\draw[color=black] (126.000000,0.000000) node[fill=white,left,minimum height=18.000000pt,minimum width=15.000000pt,inner sep=0pt] {\phantom{$\ket{0}_{A_2}$}};
\draw[color=black] (126.000000,0.000000) node[left] {$\ket{0}_{A_2}$};
\draw (118.500000,90.000000) -- (118.500000,36.000000);
\begin{scope}
\draw[fill=white] (118.500000, 63.000000) +(-45.000000:8.485281pt and 46.669048pt) -- +(45.000000:8.485281pt and 46.669048pt) -- +(135.000000:8.485281pt and 46.669048pt) -- +(225.000000:8.485281pt and 46.669048pt) -- cycle;
\clip (118.500000, 63.000000) +(-45.000000:8.485281pt and 46.669048pt) -- +(45.000000:8.485281pt and 46.669048pt) -- +(135.000000:8.485281pt and 46.669048pt) -- +(225.000000:8.485281pt and 46.669048pt) -- cycle;
\draw (118.500000, 63.000000) node {$E$};
\end{scope}
\begin{scope}
\draw[fill=white] (138.000000, 18.000000) +(-45.000000:8.485281pt and 8.485281pt) -- +(45.000000:8.485281pt and 8.485281pt) -- +(135.000000:8.485281pt and 8.485281pt) -- +(225.000000:8.485281pt and 8.485281pt) -- cycle;
\clip (138.000000, 18.000000) +(-45.000000:8.485281pt and 8.485281pt) -- +(45.000000:8.485281pt and 8.485281pt) -- +(135.000000:8.485281pt and 8.485281pt) -- +(225.000000:8.485281pt and 8.485281pt) -- cycle;
\draw (138.000000, 18.000000) node {$H$};
\end{scope}
\begin{scope}
\draw[fill=white] (138.000000, -0.000000) +(-45.000000:8.485281pt and 8.485281pt) -- +(45.000000:8.485281pt and 8.485281pt) -- +(135.000000:8.485281pt and 8.485281pt) -- +(225.000000:8.485281pt and 8.485281pt) -- cycle;
\clip (138.000000, -0.000000) +(-45.000000:8.485281pt and 8.485281pt) -- +(45.000000:8.485281pt and 8.485281pt) -- +(135.000000:8.485281pt and 8.485281pt) -- +(225.000000:8.485281pt and 8.485281pt) -- cycle;
\draw (138.000000, -0.000000) node {$H$};
\end{scope}
\draw (161.500000,90.000000) -- (161.500000,18.000000);
\begin{scope}
\draw[fill=white] (161.500000, 63.000000) +(-45.000000:16.263456pt and 46.669048pt) -- +(45.000000:16.263456pt and 46.669048pt) -- +(135.000000:16.263456pt and 46.669048pt) -- +(225.000000:16.263456pt and 46.669048pt) -- cycle;
\clip (161.500000, 63.000000) +(-45.000000:16.263456pt and 46.669048pt) -- +(45.000000:16.263456pt and 46.669048pt) -- +(135.000000:16.263456pt and 46.669048pt) -- +(225.000000:16.263456pt and 46.669048pt) -- cycle;
\draw (161.500000, 63.000000) node {$\def\arraystretch{1.7}\begin{matrix}Z_1\\Z_2\\Z_3\\Z_4\end{matrix}$};
\end{scope}
\filldraw (161.500000, 18.000000) circle(1.500000pt);
\draw (190.500000,90.000000) -- (190.500000,0.000000);
\begin{scope}
\draw[fill=white] (190.500000, 63.000000) +(-45.000000:16.263456pt and 46.669048pt) -- +(45.000000:16.263456pt and 46.669048pt) -- +(135.000000:16.263456pt and 46.669048pt) -- +(225.000000:16.263456pt and 46.669048pt) -- cycle;
\clip (190.500000, 63.000000) +(-45.000000:16.263456pt and 46.669048pt) -- +(45.000000:16.263456pt and 46.669048pt) -- +(135.000000:16.263456pt and 46.669048pt) -- +(225.000000:16.263456pt and 46.669048pt) -- cycle;
\draw (190.500000, 63.000000) node {$\def\arraystretch{1.7}\begin{matrix}X_1\\X_2\\X_3\\X_4\end{matrix}$};
\end{scope}
\filldraw (190.500000, 0.000000) circle(1.500000pt);
\begin{scope}
\draw[fill=white] (214.000000, -0.000000) +(-45.000000:8.485281pt and 8.485281pt) -- +(45.000000:8.485281pt and 8.485281pt) -- +(135.000000:8.485281pt and 8.485281pt) -- +(225.000000:8.485281pt and 8.485281pt) -- cycle;
\clip (214.000000, -0.000000) +(-45.000000:8.485281pt and 8.485281pt) -- +(45.000000:8.485281pt and 8.485281pt) -- +(135.000000:8.485281pt and 8.485281pt) -- +(225.000000:8.485281pt and 8.485281pt) -- cycle;
\draw (214.000000, -0.000000) node {$H$};
\end{scope}
\begin{scope}
\draw[fill=white] (214.000000, 18.000000) +(-45.000000:8.485281pt and 8.485281pt) -- +(45.000000:8.485281pt and 8.485281pt) -- +(135.000000:8.485281pt and 8.485281pt) -- +(225.000000:8.485281pt and 8.485281pt) -- cycle;
\clip (214.000000, 18.000000) +(-45.000000:8.485281pt and 8.485281pt) -- +(45.000000:8.485281pt and 8.485281pt) -- +(135.000000:8.485281pt and 8.485281pt) -- +(225.000000:8.485281pt and 8.485281pt) -- cycle;
\draw (214.000000, 18.000000) node {$H$};
\end{scope}
\draw[fill=white] (226.000000, 12.000000) rectangle (238.000000, 24.000000);
\draw[very thin] (232.000000, 18.600000) arc (90:150:6.000000pt);
\draw[very thin] (232.000000, 18.600000) arc (90:30:6.000000pt);
\draw[->,>=stealth] (232.000000, 12.600000) -- +(80:10.392305pt);
\draw[fill=white] (226.000000, -6.000000) rectangle (238.000000, 6.000000);
\draw[very thin] (232.000000, 0.600000) arc (90:150:6.000000pt);
\draw[very thin] (232.000000, 0.600000) arc (90:30:6.000000pt);
\draw[->,>=stealth] (232.000000, -5.400000) -- +(80:10.392305pt);
\draw[decorate,decoration={brace,amplitude = 4.000000pt},very thick] (7.500000,99.000000) -- (70.500000,99.000000);
\draw (39.000000, 103.000000) node[text width=144pt,above,text centered] {Encoder};
\draw[decorate,decoration={brace,amplitude = 4.000000pt},very thick] (130.500000,99.000000) -- (221.500000,99.000000);
\draw (176.000000, 103.000000) node[text width=144pt,above,text centered] {{Syndrome extraction}};
\end{tikzpicture}

%% file: figs/qec/qec_proc2.tikz
\usetikzlibrary{decorations.pathreplacing,decorations.pathmorphing}
\providecommand{\ket}[1]{\left|#1\right\rangle}
\begin{tikzpicture}[scale=1.500000,x=1pt,y=1pt]
\filldraw[color=white] (0.000000, -12.000000) rectangle (187.000000, 36.000000);
\draw[color=black] (0.000000,24.000000) -- (187.000000,24.000000);
\draw[color=black] (0.000000,24.000000) node[left] {$\ket{\psi}_L$};
\draw[color=black] (0.000000,0.000000) -- (67.000000,0.000000);
\draw[color=black] (67.000000,-0.500000) -- (142.000000,-0.500000);
\draw[color=black] (67.000000,0.500000) -- (142.000000,0.500000);
\draw[color=black] (0.000000,0.000000) node[left] {$\ket{A}^{\otimes m}$};
\draw (3.000000, 18.000000) -- (11.000000, 30.000000);
\draw (3.000000, -6.000000) -- (11.000000, 6.000000);
\begin{scope}
\draw[fill=white] (23.000000, 24.000000) +(-45.000000:8.485281pt and 8.485281pt) -- +(45.000000:8.485281pt and 8.485281pt) -- +(135.000000:8.485281pt and 8.485281pt) -- +(225.000000:8.485281pt and 8.485281pt) -- cycle;
\clip (23.000000, 24.000000) +(-45.000000:8.485281pt and 8.485281pt) -- +(45.000000:8.485281pt and 8.485281pt) -- +(135.000000:8.485281pt and 8.485281pt) -- +(225.000000:8.485281pt and 8.485281pt) -- cycle;
\draw (23.000000, 24.000000) node {{$E$}};
\end{scope}
\draw (45.000000,24.000000) -- (45.000000,0.000000);
\begin{scope}
\draw[fill=white] (45.000000, 24.000000) +(-45.000000:14.142136pt and 8.485281pt) -- +(45.000000:14.142136pt and 8.485281pt) -- +(135.000000:14.142136pt and 8.485281pt) -- +(225.000000:14.142136pt and 8.485281pt) -- cycle;
\clip (45.000000, 24.000000) +(-45.000000:14.142136pt and 8.485281pt) -- +(45.000000:14.142136pt and 8.485281pt) -- +(135.000000:14.142136pt and 8.485281pt) -- +(225.000000:14.142136pt and 8.485281pt) -- cycle;
\draw (45.000000, 24.000000) node {{$\mathcal{S}$}};
\end{scope}
\filldraw (45.000000, 0.000000) circle(1.500000pt);
\filldraw (67.000000, 0.000000) circle(1.500000pt);
\draw[fill=white] (61.000000, -6.000000) rectangle (73.000000, 6.000000);
\draw[very thin] (67.000000, 0.600000) arc (90:150:6.000000pt);
\draw[very thin] (67.000000, 0.600000) arc (90:30:6.000000pt);
\draw[->,>=stealth] (67.000000, -5.400000) -- +(80:10.392305pt);
\draw[fill=white,color=white] (79.000000, -6.000000) rectangle (94.000000, 6.000000);
\draw (86.500000, 0.000000) node {$S$};
\begin{scope}
\draw[fill=white] (115.000000, -0.000000) +(-45.000000:21.213203pt and 8.485281pt) -- +(45.000000:21.213203pt and 8.485281pt) -- +(135.000000:21.213203pt and 8.485281pt) -- +(225.000000:21.213203pt and 8.485281pt) -- cycle;
\clip (115.000000, -0.000000) +(-45.000000:21.213203pt and 8.485281pt) -- +(45.000000:21.213203pt and 8.485281pt) -- +(135.000000:21.213203pt and 8.485281pt) -- +(225.000000:21.213203pt and 8.485281pt) -- cycle;
\draw (115.000000, -0.000000) node {{Decoder}};
\end{scope}
\draw (141.500000,24.000000) -- (141.500000,0.000000);
\draw (142.500000,24.000000) -- (142.500000,0.000000);
\begin{scope}
\draw[fill=white] (142.000000, 24.000000) +(-45.000000:8.485281pt and 8.485281pt) -- +(45.000000:8.485281pt and 8.485281pt) -- +(135.000000:8.485281pt and 8.485281pt) -- +(225.000000:8.485281pt and 8.485281pt) -- cycle;
\clip (142.000000, 24.000000) +(-45.000000:8.485281pt and 8.485281pt) -- +(45.000000:8.485281pt and 8.485281pt) -- +(135.000000:8.485281pt and 8.485281pt) -- +(225.000000:8.485281pt and 8.485281pt) -- cycle;
\draw (142.000000, 24.000000) node {{$\mathcal{R}$}};
\end{scope}
\filldraw (142.000000, 0.000000) circle(1.500000pt);
\draw[fill=white,color=white] (154.000000, 18.000000) rectangle (184.000000, 30.000000);
\draw (169.000000, 24.000000) node {$\mathcal{R}E\ket{\psi}_L$};
\end{tikzpicture}

%% file: tikz/surface_15.tikz
\begin{tikzpicture}
\draw[red,thick] (0.000000,0.000000) -- (1.500000,0.000000);
\draw[blue,thick] (0.000000,0.000000) -- (0.000000,-1.500000);
\draw[blue,thick] (0.000000,-1.500000) -- (1.500000,-1.500000);
\draw[red,thick] (1.500000,0.000000) -- (1.500000,-1.500000);

\filldraw[fill=white, draw=black] (0.000000,0.000000) circle (0.300000) node [label=above right:{$\scriptstyle D_{1} $}]{};

\filldraw[fill=white, draw=black] (1.305000,-0.195000) rectangle (1.695000,0.195000) node [label={[label distance=-0.3cm]30:{$\scriptstyle A_{1}$}}]{};

\filldraw[fill=white, draw=black] (-0.195000,-1.695000) rectangle (0.195000,-1.305000) node [label={[label distance=-0.3cm]30:{$\scriptstyle A_{2}$}}]{};

\filldraw[fill=white, draw=black] (1.500000,-1.500000) circle (0.300000) node [label=above right:{$\scriptstyle D_{2} $}]{};
\end{tikzpicture}

%% file: tikz/four_cycle.tikz
\providecommand{\ket}[1]{\left|#1\right\rangle}
\begin{tikzpicture}[scale=1.500000,x=1pt,y=1pt]
\filldraw[color=white] (0.000000, -7.500000) rectangle (120.000000, 52.500000);
\draw[color=black] (0.000000,45.000000) -- (120.000000,45.000000);
\draw[color=black] (0.000000,45.000000) node[left] {$\ket{D_1}$};
\draw[color=black] (0.000000,30.000000) -- (120.000000,30.000000);
\draw[color=black] (0.000000,30.000000) node[left] {$\ket{D_2}$};
\draw[color=black] (0.000000,15.000000) -- (111.000000,15.000000);
\draw[color=black] (111.000000,14.500000) -- (120.000000,14.500000);
\draw[color=black] (111.000000,15.500000) -- (120.000000,15.500000);
\draw[color=black] (0.000000,15.000000) node[left] {$\ket{0}_{A1}$};
\draw[color=black] (0.000000,0.000000) -- (111.000000,0.000000);
\draw[color=black] (111.000000,-0.500000) -- (120.000000,-0.500000);
\draw[color=black] (111.000000,0.500000) -- (120.000000,0.500000);
\draw[color=black] (0.000000,0.000000) node[left] {$\ket{0}_{A2}$};
\begin{scope}
\draw[fill=white] (9.000000, 15.000000) +(-45.000000:8.485281pt and 8.485281pt) -- +(45.000000:8.485281pt and 8.485281pt) -- +(135.000000:8.485281pt and 8.485281pt) -- +(225.000000:8.485281pt and 8.485281pt) -- cycle;
\clip (9.000000, 15.000000) +(-45.000000:8.485281pt and 8.485281pt) -- +(45.000000:8.485281pt and 8.485281pt) -- +(135.000000:8.485281pt and 8.485281pt) -- +(225.000000:8.485281pt and 8.485281pt) -- cycle;
\draw (9.000000, 15.000000) node {$H$};
\end{scope}
\begin{scope}
\draw[fill=white] (9.000000, -0.000000) +(-45.000000:8.485281pt and 8.485281pt) -- +(45.000000:8.485281pt and 8.485281pt) -- +(135.000000:8.485281pt and 8.485281pt) -- +(225.000000:8.485281pt and 8.485281pt) -- cycle;
\clip (9.000000, -0.000000) +(-45.000000:8.485281pt and 8.485281pt) -- +(45.000000:8.485281pt and 8.485281pt) -- +(135.000000:8.485281pt and 8.485281pt) -- +(225.000000:8.485281pt and 8.485281pt) -- cycle;
\draw (9.000000, -0.000000) node {$H$};
\end{scope}
\draw (27.000000,45.000000) -- (27.000000,15.000000);
\begin{scope}
\draw[fill=white] (27.000000, 45.000000) +(-45.000000:8.485281pt and 8.485281pt) -- +(45.000000:8.485281pt and 8.485281pt) -- +(135.000000:8.485281pt and 8.485281pt) -- +(225.000000:8.485281pt and 8.485281pt) -- cycle;
\clip (27.000000, 45.000000) +(-45.000000:8.485281pt and 8.485281pt) -- +(45.000000:8.485281pt and 8.485281pt) -- +(135.000000:8.485281pt and 8.485281pt) -- +(225.000000:8.485281pt and 8.485281pt) -- cycle;
\draw (27.000000, 45.000000) node {$X$};
\end{scope}
\filldraw (27.000000, 15.000000) circle(1.500000pt);
\draw (45.000000,30.000000) -- (45.000000,15.000000);
\begin{scope}
\draw[fill=white] (45.000000, 30.000000) +(-45.000000:8.485281pt and 8.485281pt) -- +(45.000000:8.485281pt and 8.485281pt) -- +(135.000000:8.485281pt and 8.485281pt) -- +(225.000000:8.485281pt and 8.485281pt) -- cycle;
\clip (45.000000, 30.000000) +(-45.000000:8.485281pt and 8.485281pt) -- +(45.000000:8.485281pt and 8.485281pt) -- +(135.000000:8.485281pt and 8.485281pt) -- +(225.000000:8.485281pt and 8.485281pt) -- cycle;
\draw (45.000000, 30.000000) node {$X$};
\end{scope}
\filldraw (45.000000, 15.000000) circle(1.500000pt);
\draw (57.000000,45.000000) -- (57.000000,0.000000);
\begin{scope}
\draw[fill=white] (57.000000, 45.000000) +(-45.000000:8.485281pt and 8.485281pt) -- +(45.000000:8.485281pt and 8.485281pt) -- +(135.000000:8.485281pt and 8.485281pt) -- +(225.000000:8.485281pt and 8.485281pt) -- cycle;
\clip (57.000000, 45.000000) +(-45.000000:8.485281pt and 8.485281pt) -- +(45.000000:8.485281pt and 8.485281pt) -- +(135.000000:8.485281pt and 8.485281pt) -- +(225.000000:8.485281pt and 8.485281pt) -- cycle;
\draw (57.000000, 45.000000) node {$Z$};
\end{scope}
\filldraw (57.000000, 0.000000) circle(1.500000pt);
\draw (75.000000,30.000000) -- (75.000000,0.000000);
\begin{scope}
\draw[fill=white] (75.000000, 30.000000) +(-45.000000:8.485281pt and 8.485281pt) -- +(45.000000:8.485281pt and 8.485281pt) -- +(135.000000:8.485281pt and 8.485281pt) -- +(225.000000:8.485281pt and 8.485281pt) -- cycle;
\clip (75.000000, 30.000000) +(-45.000000:8.485281pt and 8.485281pt) -- +(45.000000:8.485281pt and 8.485281pt) -- +(135.000000:8.485281pt and 8.485281pt) -- +(225.000000:8.485281pt and 8.485281pt) -- cycle;
\draw (75.000000, 30.000000) node {$Z$};
\end{scope}
\filldraw (75.000000, 0.000000) circle(1.500000pt);
\begin{scope}
\draw[fill=white] (93.000000, 15.000000) +(-45.000000:8.485281pt and 8.485281pt) -- +(45.000000:8.485281pt and 8.485281pt) -- +(135.000000:8.485281pt and 8.485281pt) -- +(225.000000:8.485281pt and 8.485281pt) -- cycle;
\clip (93.000000, 15.000000) +(-45.000000:8.485281pt and 8.485281pt) -- +(45.000000:8.485281pt and 8.485281pt) -- +(135.000000:8.485281pt and 8.485281pt) -- +(225.000000:8.485281pt and 8.485281pt) -- cycle;
\draw (93.000000, 15.000000) node {$H$};
\end{scope}
\begin{scope}
\draw[fill=white] (93.000000, -0.000000) +(-45.000000:8.485281pt and 8.485281pt) -- +(45.000000:8.485281pt and 8.485281pt) -- +(135.000000:8.485281pt and 8.485281pt) -- +(225.000000:8.485281pt and 8.485281pt) -- cycle;
\clip (93.000000, -0.000000) +(-45.000000:8.485281pt and 8.485281pt) -- +(45.000000:8.485281pt and 8.485281pt) -- +(135.000000:8.485281pt and 8.485281pt) -- +(225.000000:8.485281pt and 8.485281pt) -- cycle;
\draw (93.000000, -0.000000) node {$H$};
\end{scope}
\draw[fill=white] (105.000000, 9.000000) rectangle (117.000000, 21.000000);
\draw[very thin] (111.000000, 15.600000) arc (90:150:6.000000pt);
\draw[very thin] (111.000000, 15.600000) arc (90:30:6.000000pt);
\draw[->,>=stealth] (111.000000, 9.600000) -- +(80:10.392305pt);
\draw[fill=white] (105.000000, -6.000000) rectangle (117.000000, 6.000000);
\draw[very thin] (111.000000, 0.600000) arc (90:150:6.000000pt);
\draw[very thin] (111.000000, 0.600000) arc (90:30:6.000000pt);
\draw[->,>=stealth] (111.000000, -5.400000) -- +(80:10.392305pt);
\end{tikzpicture}

%% file: tikz/surface_2.tikz
\begin{tikzpicture}
\draw[red,thick] (0.000000,0.000000) -- (3.000000,0.000000);
\draw[blue,thick] (0.000000,0.000000) -- (0.000000,-3.000000);
\draw[blue,thick] (0.000000,-1.500000) -- (3.000000,-1.500000);
\draw[red,thick] (1.500000,0.000000) -- (1.500000,-3.000000);
\draw[red,thick] (0.000000,-3.000000) -- (3.000000,-3.000000);
\draw[blue,thick] (3.000000,0.000000) -- (3.000000,-3.000000);

\filldraw[fill=white, draw=black] (0.000000,0.000000) circle (0.300000) node [label=above right:{$\scriptstyle D_{1} $}]{};

\filldraw[fill=white, draw=black] (1.305000,-0.195000) rectangle (1.695000,0.195000) node [label={[label distance=-0.3cm]30:{$\scriptstyle A_{1}$}}]{};

\filldraw[fill=white, draw=black] (3.000000,0.000000) circle (0.300000) node [label=above right:{$\scriptstyle D_{2} $}]{};

\filldraw[fill=white, draw=black] (-0.195000,-1.695000) rectangle (0.195000,-1.305000) node [label={[label distance=-0.3cm]30:{$\scriptstyle A_{2}$}}]{};

\filldraw[fill=white, draw=black] (1.500000,-1.500000) circle (0.300000) node [label=above right:{$\scriptstyle D_{3} $}]{};

\filldraw[fill=white, draw=black] (2.805000,-1.695000) rectangle (3.195000,-1.305000) node [label={[label distance=-0.3cm]30:{$\scriptstyle A_{3}$}}]{};

\filldraw[fill=white, draw=black] (0.000000,-3.000000) circle (0.300000) node [label=above right:{$\scriptstyle D_{4} $}]{};

\filldraw[fill=white, draw=black] (1.305000,-3.195000) rectangle (1.695000,-2.805000) node [label={[label distance=-0.3cm]30:{$\scriptstyle A_{4}$}}]{};

\filldraw[fill=white, draw=black] (3.000000,-3.000000) circle (0.300000) node [label=above right:{$\scriptstyle D_{5} $}]{};
\end{tikzpicture}

%% file: tikz/surface_2_error.tikz
\begin{tikzpicture}
\draw[red,thick] (0.000000,0.000000) -- (3.000000,0.000000);
\draw[blue,thick] (0.000000,0.000000) -- (0.000000,-3.000000);
\draw[blue,thick] (0.000000,-1.500000) -- (3.000000,-1.500000);
\draw[red,thick] (1.500000,0.000000) -- (1.500000,-3.000000);
\draw[red,thick] (0.000000,-3.000000) -- (3.000000,-3.000000);
\draw[blue,thick] (3.000000,0.000000) -- (3.000000,-3.000000);

\filldraw[fill=white, draw=black] (0.000000,0.000000) circle (0.300000) node [label=above right:{$\scriptstyle D_{1} $}]{Z};

\filldraw[fill=red, draw=black] (1.305000,-0.195000) rectangle (1.695000,0.195000) node [label={[label distance=-0.3cm]30:{$\scriptstyle A_{1}$}}]{};

\filldraw[fill=white, draw=black] (3.000000,0.000000) circle (0.300000) node [label=above right:{$\scriptstyle D_{2} $}]{};

\filldraw[fill=white, draw=black] (-0.195000,-1.695000) rectangle (0.195000,-1.305000) node [label={[label distance=-0.3cm]30:{$\scriptstyle A_{2}$}}]{};

\filldraw[fill=white, draw=black] (1.500000,-1.500000) circle (0.300000) node [label=above right:{$\scriptstyle D_{3} $}]{};

\filldraw[fill=red, draw=black] (2.805000,-1.695000) rectangle (3.195000,-1.305000) node [label={[label distance=-0.3cm]30:{$\scriptstyle A_{3}$}}]{};

\filldraw[fill=white, draw=black] (0.000000,-3.000000) circle (0.300000) node [label=above right:{$\scriptstyle D_{4} $}]{};

\filldraw[fill=white, draw=black] (1.305000,-3.195000) rectangle (1.695000,-2.805000) node [label={[label distance=-0.3cm]30:{$\scriptstyle A_{4}$}}]{};

\filldraw[fill=white, draw=black] (3.000000,-3.000000) circle (0.300000) node [label=above right:{$\scriptstyle D_{5} $}]{X};
\end{tikzpicture}

%% file: tikz/surface_2_lx.tikz
\begin{tikzpicture}
\draw[red,thick] (0.000000,0.000000) -- (3.000000,0.000000);
\draw[blue,thick] (0.000000,0.000000) -- (0.000000,-3.000000);
\draw[blue,thick] (0.000000,-1.500000) -- (3.000000,-1.500000);
\draw[red,thick] (1.500000,0.000000) -- (1.500000,-3.000000);
\draw[red,thick] (0.000000,-3.000000) -- (3.000000,-3.000000);
\draw[blue,thick] (3.000000,0.000000) -- (3.000000,-3.000000);

\filldraw[fill=white, draw=black] (0.000000,0.000000) circle (0.300000) node [label=above right:{$\scriptstyle D_{1} $}]{X};

\filldraw[fill=white, draw=black] (1.305000,-0.195000) rectangle (1.695000,0.195000) node [label={[label distance=-0.3cm]30:{$\scriptstyle A_{1}$}}]{};

\filldraw[fill=white, draw=black] (3.000000,0.000000) circle (0.300000) node [label=above right:{$\scriptstyle D_{2} $}]{};

\filldraw[fill=white, draw=black] (-0.195000,-1.695000) rectangle (0.195000,-1.305000) node [label={[label distance=-0.3cm]30:{$\scriptstyle A_{2}$}}]{};

\filldraw[fill=white, draw=black] (1.500000,-1.500000) circle (0.300000) node [label=above right:{$\scriptstyle D_{3} $}]{};

\filldraw[fill=white, draw=black] (2.805000,-1.695000) rectangle (3.195000,-1.305000) node [label={[label distance=-0.3cm]30:{$\scriptstyle A_{3}$}}]{};

\filldraw[fill=white, draw=black] (0.000000,-3.000000) circle (0.300000) node [label=above right:{$\scriptstyle D_{4} $}]{X};

\filldraw[fill=white, draw=black] (1.305000,-3.195000) rectangle (1.695000,-2.805000) node [label={[label distance=-0.3cm]30:{$\scriptstyle A_{4}$}}]{};

\filldraw[fill=white, draw=black] (3.000000,-3.000000) circle (0.300000) node [label=above right:{$\scriptstyle D_{5} $}]{};
\end{tikzpicture}

%% file: tikz/surface_2_lz.tikz
\begin{tikzpicture}
\draw[red,thick] (0.000000,0.000000) -- (3.000000,0.000000);
\draw[blue,thick] (0.000000,0.000000) -- (0.000000,-3.000000);
\draw[blue,thick] (0.000000,-1.500000) -- (3.000000,-1.500000);
\draw[red,thick] (1.500000,0.000000) -- (1.500000,-3.000000);
\draw[red,thick] (0.000000,-3.000000) -- (3.000000,-3.000000);
\draw[blue,thick] (3.000000,0.000000) -- (3.000000,-3.000000);

\filldraw[fill=white, draw=black] (0.000000,0.000000) circle (0.300000) node [label=above right:{$\scriptstyle D_{1} $}]{Z};

\filldraw[fill=white, draw=black] (1.305000,-0.195000) rectangle (1.695000,0.195000) node [label={[label distance=-0.3cm]30:{$\scriptstyle A_{1}$}}]{};

\filldraw[fill=white, draw=black] (3.000000,0.000000) circle (0.300000) node [label=above right:{$\scriptstyle D_{2} $}]{Z};

\filldraw[fill=white, draw=black] (-0.195000,-1.695000) rectangle (0.195000,-1.305000) node [label={[label distance=-0.3cm]30:{$\scriptstyle A_{2}$}}]{};

\filldraw[fill=white, draw=black] (1.500000,-1.500000) circle (0.300000) node [label=above right:{$\scriptstyle D_{3} $}]{};

\filldraw[fill=white, draw=black] (2.805000,-1.695000) rectangle (3.195000,-1.305000) node [label={[label distance=-0.3cm]30:{$\scriptstyle A_{3}$}}]{};

\filldraw[fill=white, draw=black] (0.000000,-3.000000) circle (0.300000) node [label=above right:{$\scriptstyle D_{4} $}]{};

\filldraw[fill=white, draw=black] (1.305000,-3.195000) rectangle (1.695000,-2.805000) node [label={[label distance=-0.3cm]30:{$\scriptstyle A_{4}$}}]{};

\filldraw[fill=white, draw=black] (3.000000,-3.000000) circle (0.300000) node [label=above right:{$\scriptstyle D_{5} $}]{};
\end{tikzpicture}

%% file: tikz/surface_3.tikz
\begin{tikzpicture}
\draw[red,thick] (0.000000,0.000000) -- (6.000000,0.000000);
\draw[blue,thick] (0.000000,0.000000) -- (0.000000,-6.000000);
\draw[blue,thick] (0.000000,-1.500000) -- (6.000000,-1.500000);
\draw[red,thick] (1.500000,0.000000) -- (1.500000,-6.000000);
\draw[red,thick] (0.000000,-3.000000) -- (6.000000,-3.000000);
\draw[blue,thick] (3.000000,0.000000) -- (3.000000,-6.000000);
\draw[blue,thick] (0.000000,-4.500000) -- (6.000000,-4.500000);
\draw[red,thick] (4.500000,0.000000) -- (4.500000,-6.000000);
\draw[red,thick] (0.000000,-6.000000) -- (6.000000,-6.000000);
\draw[blue,thick] (6.000000,0.000000) -- (6.000000,-6.000000);

\filldraw[fill=white, draw=black] (0.000000,0.000000) circle (0.300000) node [label=above right:{$\scriptstyle D_{1} $}]{};

\filldraw[fill=white, draw=black] (1.305000,-0.195000) rectangle (1.695000,0.195000) node [label={[label distance=-0.3cm]30:{$\scriptstyle A_{1}$}}]{};

\filldraw[fill=white, draw=black] (3.000000,0.000000) circle (0.300000) node [label=above right:{$\scriptstyle D_{2} $}]{};

\filldraw[fill=white, draw=black] (4.305000,-0.195000) rectangle (4.695000,0.195000) node [label={[label distance=-0.3cm]30:{$\scriptstyle A_{2}$}}]{};

\filldraw[fill=white, draw=black] (6.000000,0.000000) circle (0.300000) node [label=above right:{$\scriptstyle D_{3} $}]{};

\filldraw[fill=white, draw=black] (-0.195000,-1.695000) rectangle (0.195000,-1.305000) node [label={[label distance=-0.3cm]30:{$\scriptstyle A_{3}$}}]{};

\filldraw[fill=white, draw=black] (1.500000,-1.500000) circle (0.300000) node [label=above right:{$\scriptstyle D_{4} $}]{};

\filldraw[fill=white, draw=black] (2.805000,-1.695000) rectangle (3.195000,-1.305000) node [label={[label distance=-0.3cm]30:{$\scriptstyle A_{4}$}}]{};

\filldraw[fill=white, draw=black] (4.500000,-1.500000) circle (0.300000) node [label=above right:{$\scriptstyle D_{5} $}]{};

\filldraw[fill=white, draw=black] (5.805000,-1.695000) rectangle (6.195000,-1.305000) node [label={[label distance=-0.3cm]30:{$\scriptstyle A_{5}$}}]{};

\filldraw[fill=white, draw=black] (0.000000,-3.000000) circle (0.300000) node [label=above right:{$\scriptstyle D_{6} $}]{};

\filldraw[fill=white, draw=black] (1.305000,-3.195000) rectangle (1.695000,-2.805000) node [label={[label distance=-0.3cm]30:{$\scriptstyle A_{6}$}}]{};

\filldraw[fill=white, draw=black] (3.000000,-3.000000) circle (0.300000) node [label=above right:{$\scriptstyle D_{7} $}]{};

\filldraw[fill=white, draw=black] (4.305000,-3.195000) rectangle (4.695000,-2.805000) node [label={[label distance=-0.3cm]30:{$\scriptstyle A_{7}$}}]{};

\filldraw[fill=white, draw=black] (6.000000,-3.000000) circle (0.300000) node [label=above right:{$\scriptstyle D_{8} $}]{};

\filldraw[fill=white, draw=black] (-0.195000,-4.695000) rectangle (0.195000,-4.305000) node [label={[label distance=-0.3cm]30:{$\scriptstyle A_{8}$}}]{};

\filldraw[fill=white, draw=black] (1.500000,-4.500000) circle (0.300000) node [label=above right:{$\scriptstyle D_{9} $}]{};

\filldraw[fill=white, draw=black] (2.805000,-4.695000) rectangle (3.195000,-4.305000) node [label={[label distance=-0.3cm]30:{$\scriptstyle A_{9}$}}]{};

\filldraw[fill=white, draw=black] (4.500000,-4.500000) circle (0.300000) node [label=above right:{$\scriptstyle D_{10} $}]{};

\filldraw[fill=white, draw=black] (5.805000,-4.695000) rectangle (6.195000,-4.305000) node [label={[label distance=-0.3cm]30:{$\scriptstyle A_{10}$}}]{};

\filldraw[fill=white, draw=black] (0.000000,-6.000000) circle (0.300000) node [label=above right:{$\scriptstyle D_{11} $}]{};

\filldraw[fill=white, draw=black] (1.305000,-6.195000) rectangle (1.695000,-5.805000) node [label={[label distance=-0.3cm]30:{$\scriptstyle A_{11}$}}]{};

\filldraw[fill=white, draw=black] (3.000000,-6.000000) circle (0.300000) node [label=above right:{$\scriptstyle D_{12} $}]{};

\filldraw[fill=white, draw=black] (4.305000,-6.195000) rectangle (4.695000,-5.805000) node [label={[label distance=-0.3cm]30:{$\scriptstyle A_{12}$}}]{};

\filldraw[fill=white, draw=black] (6.000000,-6.000000) circle (0.300000) node [label=above right:{$\scriptstyle D_{13} $}]{};
\end{tikzpicture}

%% file: figs/qec/xz.tikz
\providecommand{\ket}[1]{\left|#1\right\rangle}
\begin{tikzpicture}[scale=1.500000,x=1pt,y=1pt]
\filldraw[color=white] (0.000000, -9.000000) rectangle (48.000000, 9.000000);
\draw[color=black] (0.000000,0.000000) -- (48.000000,0.000000);
\draw[color=black] (0.000000,0.000000) node[left] {$\ket{\psi}$};
\begin{scope}
\draw[fill=white] (12.000000, -0.000000) +(-45.000000:8.485281pt and 8.485281pt) -- +(45.000000:8.485281pt and 8.485281pt) -- +(135.000000:8.485281pt and 8.485281pt) -- +(225.000000:8.485281pt and 8.485281pt) -- cycle;
\clip (12.000000, -0.000000) +(-45.000000:8.485281pt and 8.485281pt) -- +(45.000000:8.485281pt and 8.485281pt) -- +(135.000000:8.485281pt and 8.485281pt) -- +(225.000000:8.485281pt and 8.485281pt) -- cycle;
\draw (12.000000, -0.000000) node {$Z$};
\end{scope}
\begin{scope}
\draw[fill=white] (36.000000, -0.000000) +(-45.000000:8.485281pt and 8.485281pt) -- +(45.000000:8.485281pt and 8.485281pt) -- +(135.000000:8.485281pt and 8.485281pt) -- +(225.000000:8.485281pt and 8.485281pt) -- cycle;
\clip (36.000000, -0.000000) +(-45.000000:8.485281pt and 8.485281pt) -- +(45.000000:8.485281pt and 8.485281pt) -- +(135.000000:8.485281pt and 8.485281pt) -- +(225.000000:8.485281pt and 8.485281pt) -- cycle;
\draw (36.000000, -0.000000) node {$X$};
\end{scope}
\draw[color=black] (48.000000,0.000000) node[right] {$XZ\ket{\psi}$};
\end{tikzpicture}

%% file: figs/qec/multi.tikz
\providecommand{\ket}[1]{\left|#1\right\rangle}
\begin{tikzpicture}[scale=1.500000,x=1pt,y=1pt]
\filldraw[color=white] (0.000000, -9.000000) rectangle (44.000000, 27.000000);
\draw[color=black] (0.000000,18.000000) -- (44.000000,18.000000);
\draw[color=black] (0.000000,18.000000) node[left] {$\ket{0}$};
\draw[color=black] (0.000000,0.000000) -- (44.000000,0.000000);
\draw[color=black] (0.000000,0.000000) node[left] {$\ket{0}$};
\draw (22.000000,18.000000) -- (22.000000,0.000000);
\begin{scope}
\draw[fill=white] (22.000000, 9.000000) +(-45.000000:14.142136pt and 21.213203pt) -- +(45.000000:14.142136pt and 21.213203pt) -- +(135.000000:14.142136pt and 21.213203pt) -- +(225.000000:14.142136pt and 21.213203pt) -- cycle;
\clip (22.000000, 9.000000) +(-45.000000:14.142136pt and 21.213203pt) -- +(45.000000:14.142136pt and 21.213203pt) -- +(135.000000:14.142136pt and 21.213203pt) -- +(225.000000:14.142136pt and 21.213203pt) -- cycle;
\draw (22.000000, 9.000000) node {$X_1X_2$};
\end{scope}
\draw[color=black] (44.000000,18.000000) node[right] {$\ket{1}$};
\draw[color=black] (44.000000,0.000000) node[right] {$\ket{1}$};
\end{tikzpicture}

%% file: figs/qec/control.tikz
\providecommand{\ket}[1]{\left|#1\right\rangle}
\begin{tikzpicture}[scale=1.500000,x=1pt,y=1pt]
\filldraw[color=white] (0.000000, -9.000000) rectangle (36.000000, 27.000000);
\draw[color=black] (0.000000,18.000000) -- (36.000000,18.000000);
\draw[color=black] (0.000000,18.000000) node[left] {$\ket{C}$};
\draw[color=black] (0.000000,0.000000) -- (36.000000,0.000000);
\draw[color=black] (0.000000,0.000000) node[left] {$\ket{T}$};
\draw (18.000000,18.000000) -- (18.000000,0.000000);
\begin{scope}
\draw[fill=white] (18.000000, -0.000000) +(-45.000000:8.485281pt and 8.485281pt) -- +(45.000000:8.485281pt and 8.485281pt) -- +(135.000000:8.485281pt and 8.485281pt) -- +(225.000000:8.485281pt and 8.485281pt) -- cycle;
\clip (18.000000, -0.000000) +(-45.000000:8.485281pt and 8.485281pt) -- +(45.000000:8.485281pt and 8.485281pt) -- +(135.000000:8.485281pt and 8.485281pt) -- +(225.000000:8.485281pt and 8.485281pt) -- cycle;
\draw (18.000000, -0.000000) node {$G$};
\end{scope}
\filldraw (18.000000, 18.000000) circle(1.500000pt);
\end{tikzpicture}

%% file: figs/qec/cnot_equiv.tikz
\providecommand{\ket}[1]{\left|#1\right\rangle}
\begin{tikzpicture}[scale=1.500000,x=1pt,y=1pt]
\filldraw[color=white] (0.000000, -9.000000) rectangle (105.000000, 27.000000);
\draw[color=black] (0.000000,18.000000) -- (105.000000,18.000000);
\draw[color=black] (0.000000,0.000000) -- (105.000000,0.000000);
\draw (15.000000,18.000000) -- (15.000000,0.000000);
\filldraw (15.000000, 18.000000) circle(1.500000pt);
\begin{scope}
\draw[fill=white] (15.000000, 0.000000) circle(3.000000pt);
\clip (15.000000, 0.000000) circle(3.000000pt);
\draw (12.000000, 0.000000) -- (18.000000, 0.000000);
\draw (15.000000, -3.000000) -- (15.000000, 3.000000);
\end{scope}
\draw[fill=white,color=white] (42.000000, -6.000000) rectangle (57.000000, 24.000000);
\draw (49.500000, 9.000000) node {$=$};
\draw (87.000000,18.000000) -- (87.000000,0.000000);
\begin{scope}
\draw[fill=white] (87.000000, -0.000000) +(-45.000000:8.485281pt and 8.485281pt) -- +(45.000000:8.485281pt and 8.485281pt) -- +(135.000000:8.485281pt and 8.485281pt) -- +(225.000000:8.485281pt and 8.485281pt) -- cycle;
\clip (87.000000, -0.000000) +(-45.000000:8.485281pt and 8.485281pt) -- +(45.000000:8.485281pt and 8.485281pt) -- +(135.000000:8.485281pt and 8.485281pt) -- +(225.000000:8.485281pt and 8.485281pt) -- cycle;
\draw (87.000000, -0.000000) node {$X$};
\end{scope}
\filldraw (87.000000, 18.000000) circle(1.500000pt);
\end{tikzpicture}

%% file: figs/qec/hello_world.tikz
\providecommand{\ket}[1]{\left|#1\right\rangle}
\begin{tikzpicture}[scale=1.500000,x=1pt,y=1pt]
\filldraw[color=white] (0.000000, -9.000000) rectangle (72.000000, 9.000000);
\draw[color=black] (0.000000,0.000000) -- (54.000000,0.000000);
\draw[color=black] (54.000000,-0.500000) -- (72.000000,-0.500000);
\draw[color=black] (54.000000,0.500000) -- (72.000000,0.500000);
\draw[color=black] (0.000000,0.000000) node[left] {$\ket{0}$};
\begin{scope}
\draw[fill=white] (18.000000, -0.000000) +(-45.000000:8.485281pt and 8.485281pt) -- +(45.000000:8.485281pt and 8.485281pt) -- +(135.000000:8.485281pt and 8.485281pt) -- +(225.000000:8.485281pt and 8.485281pt) -- cycle;
\clip (18.000000, -0.000000) +(-45.000000:8.485281pt and 8.485281pt) -- +(45.000000:8.485281pt and 8.485281pt) -- +(135.000000:8.485281pt and 8.485281pt) -- +(225.000000:8.485281pt and 8.485281pt) -- cycle;
\draw (18.000000, -0.000000) node {$H$};
\end{scope}
\draw[fill=white] (48.000000, -6.000000) rectangle (60.000000, 6.000000);
\draw[very thin] (54.000000, 0.600000) arc (90:150:6.000000pt);
\draw[very thin] (54.000000, 0.600000) arc (90:30:6.000000pt);
\draw[->,>=stealth] (54.000000, -5.400000) -- +(80:10.392305pt);
\end{tikzpicture}